\documentclass[12pt,hidelinks]{article}
\usepackage{bm}
\usepackage[semicolon]{natbib}
\usepackage[pdftex]{graphicx}
\DeclareGraphicsExtensions{.png,.pdf,.jpg}
\usepackage[hmargin=1in,vmargin=1in]{geometry}
\usepackage{setspace}
\usepackage{verbatim}
\usepackage{amssymb, amsthm}
\usepackage{amsmath}
\usepackage{booktabs}
\usepackage{hyperref}
\usepackage{appendix}
\usepackage{tcolorbox}
\usepackage{xcolor}
\usepackage{longtable}
\usepackage{afterpage}
\usepackage{pdflscape}
\usepackage{capt-of}
\usepackage[normalem]{ulem}
\usepackage{lscape}
\usepackage{subcaption}
\usepackage{mathtools}
\usepackage[maxfloats=25]{morefloats}
\usepackage[export]{adjustbox}

\usepackage{float}

\theoremstyle{remark} 
\usepackage{tikz}
\usetikzlibrary{fit,positioning}

\newif\ifblinded

\title{Bayesian Functional Graphical Models with Change-Point Detection}

\ifblinded
\author{}
\else
\author{Chunshan Liu\thanks{Lutron, Austin, TX}, Daniel R. Kowal\thanks{Department of Statistics, Rice University, Houston, TX 77005, USA.}, James Doss-Gollin\thanks{Department of Civil and Environmental Engineering, Rice University, Houston, TX 77005, USA.} and Marina Vannucci$^\dag$\footnote{Corresponding author: marina@rice.edu}}
\fi

\begin{document}

\maketitle

\begin{abstract}

Functional data analysis, which models  data as realizations of random functions over a continuum, has emerged as a useful tool for time series data. Often, the goal is to infer the dynamic connections (or time-varying  conditional dependencies) among multiple functions or time series. For this task, a dynamic and Bayesian functional graphical model is introduced. The proposed modeling approach prioritizes the careful definition of an appropriate graph to identify both time-invariant and time-varying connectivity patterns. A novel block-structured sparsity prior is paired with a finite basis expansion, which together yield effective shrinkage and graph selection with efficient computations via a Gibbs sampling algorithm. Crucially, the model includes (one or more) graph changepoints, which are learned jointly with all model parameters and incorporate graph dynamics. Simulation studies demonstrate excellent graph selection capabilities, with significant improvements over competing methods. The proposed approach is applied to study of dynamic connectivity patterns of sea surface temperatures in the Pacific Ocean and reveals meaningful edges.

\end{abstract}

\section{Introduction}
\label{sec::introduction}
Undirected graphical models are widely used to learn the conditional dependencies among random variables. Such dependencies can be represented by a graph $\mathcal{G(V, E)}$, with vertices $\mathcal{V} = \{1,\ldots,p\}$ representing $p$ variables, and undirected edges $\mathcal{E} \subset \mathcal{V} \times \mathcal{V}$ connecting pairs of vertices. If two variables are not connected by an edge, it indicates that the two variables are conditionally independent given the remaining $p-2$ variables. 
Graphical models often focus on the scenarios where many variables are conditionally independent, resulting in sparse graphs with very few edges. These models are particularly appealing because they exclude spurious dependencies, and thus offer more reliable insights into the underlying connectivity patterns.

Traditional graphical model approaches typically apply to multivariate data, which consist of $p$ random variables observed at $n$ independently and identically distributed (iid) experimental units. Recent work has focused on extensions to more complicated scenarios. Here, we consider \emph{multivariate functional data}, as $n$ iid observations on $p$ random functions defined over a continuum such as time or space.
Multivariate functional data arises in many fields, including neuroimaging \citep{qiao2019functional}, traffic tracking \citep{wang2016functional}, and spectrometric data analysis \citep{zhu2010bayesian}, among many others. In our motivating example, the data consists of daily sea surface temperatures (SSTs) from $p = 15$ locations in the Pacific Ocean, which are not iid due to strong temporal auto-correlation.
By modeling these data as \emph{functions} of time-of-year, we seek to infer how the sea surface temperatures at different locations influence each other, and to assess whether these connectivity patterns are dynamic.

The main tool for inferring conditional dependencies among random functions is \emph{functional graphical models} (FGMs), where each node represents an entire function instead of a scalar variable. Like for traditional (non-functional) graphical models, there are primarily two approaches for graph estimation with FGMs. The first approach is neighborhood selection, which regresses each variable on all other variables and identifies conditional dependencies based on nonzero regression coefficients. \cite{zapata2019partial} and  \cite{zhao2021high}   adapt this approach for functional data using  function-on-function regressions, i.e., linear regression models with a functional response and functional covariates. Related, \cite{wu2022monitoring} estimate multiple graphs simultaneously and add a Frobenius norm penalty to enforce graph similarity. Nonparametric variations  that relax linearity and distributional assumptions (see below) are also available  \citep{li2018nonparametric,solea2021nonparametric,solea2022copula,lee2022nonparametric} and may incorporate external covariates \citep{lee2021conditional}. However, these methods are exclusively non-Bayesian, do not provide uncertainty quantification, and typically require the selection of multiple tuning parameters.

The second approach constructs a multivariate Gaussian process model and defines a graph based on the conditional dependencies among these functions. 
\cite{zhu2016bayesian} and \cite{qiao2019functional} apply finite basis expansions with Gaussian assumptions for the basis coefficients, and demonstrate that the conditional dependencies among the basis coefficients may be used to infer the conditional dependencies of the functions. In the presence of group information among the functions, \cite{moysidis2021joint} incorporate regularization toward a common graph, while \cite{zhao2019direct} directly estimate the difference between two graphs. 
Among the very few Bayesian approaches, \cite{zhu2016bayesian} establish foundational principles for Bayesian FGMs, but rely on a restrictive hyper-inverse-Wishart prior for the key covariance matrix. 

In general, these methods only estimate a single, time-invariant graph for the multivariate functional data. However,  when the data are functions of \emph{time}, it is often of interest to estimate a \emph{dynamic} graph, so that the conditional dependencies may change through time. For instance, connectivity patterns among SSTs may differ depending on the phase of the seasonal cycle, motivating dynamic graph estimation for multivariate functional data. This generalization requires careful definition of an appropriate graph. 
\cite{qiao2020doubly} propose dynamic graph estimation by (i) smoothing the functional data and (ii) separately estimating sparse graphs at each time $t$. Nodes in this dynamic graph at time $t$ represent functions evaluated \emph{only} at time $t$. 
However, the graph at time $t$ is influenced by the functional data observed on the \emph{whole} time domain as a result of the smoothing in (i), which confuses the interpretation of the dynamic graph and may introduce sensitivity to data far from $t$. Further, this approach requires selection of multiple tuning parameters, including a sparsity parameter for every time $t$, and does not provide uncertainty quantification. 
Alternatively,  \cite{zhang2021bayesian} introduce a Bayesian model that defines an edge when two functions are  conditionally dependent at time $t$ given all other functions at \emph{all other} times. Clearly, this graph definition deviates from traditional and functional graphical models: the conditional dependence here does not  correspond exactly to the nodes in the graph. Further, \cite{zhang2021bayesian} require local, basis coefficient-specific shrinkage priors, which can induce unnecessary variability unless additional structure is imposed \citep{gao2022bayesian}. 

Some of the existing dynamic graphical models (DGM) \citep{warnick2018bayesian, liu2022dynamic, franzolini2024change} focus on the detection of changepoints over time. However, there are major differences between the use cases of the dynamic functional graphical models and the DGMs. First, the data of the former are replicated realizations of $p$ functions that are defined on a time interval, while the data of the latter are a single realization of multivariate time series. Second, dynamic functional graphical models are used when the data can be treated as smooth curves over time, while the data for DGMs are usually much noisier than smooth curves. \cite{liu2022dynamic} provide a thorough summary of existing DGMs.

In this paper, we introduce a Bayesian model and posterior sampling algorithm for dynamic FGMs. Like nearly all previous work, we construct a finite basis expansion for each function and, by assuming Gaussianity, enable graph selection via sparse estimation of the precision matrix of the basis coefficients. Our approach has two novel components. First, we specify a block-structured sparsity prior for this precision matrix, which simultaneously delivers  sparse estimates of the precision matrix
in the functional space while sharing information among the coefficients in the basis space. By design, this prior enables sparsity, smoothness, and efficient Markov chain Monte Carlo (MCMC) sampling. Second, we include (one or more)  changepoints to incorporate graph dynamics. This strategy partitions the functional data domain to infer local, time-invariant 
graphs---each of which retains the traditional FGM interpretation on its respective subdomain. As with (non-dynamic) FGMs, each node represents a random function, and edges are defined via conditional dependencies  among nodes \emph{within} the graph. 
The main difference compared to existing dynamic FGMs is that our changepoints  \emph{localize} (in time) these graphs so that they are not dependent on data observed far away (in time). Thus, we infer both time-invariant and time-varying graphs while preserving the interpretability of graphical models.

We apply our proposed approach to analyze the spatiotemporal structure of sea surface temperatures (SSTs) in the Pacific Ocean. Large-scale circulations of the ocean vary at multiple spatial and scales, including the diurnal cycle and annual cycles as well as quasi-cyclical modes of variability ranging from a few weeks to a few years. A notable such mechanism is the El Niño-Southern Oscillation (ENSO), which describes the leading mode of internal variability of the tropical Pacific, and which affects weather globally. Warmer water in the tropical Pacific leads to increased deep convection, which can create waves in the atmosphere; thus, SSTs at two distant points may exhibit substantial covariance because they are connected by an atmospheric ``bridge'' or ``teleconnection'' \citep{Zebiak:1987cl,Angstrom:1935ej}.
These patterns often exhibit regime-like behavior which has been modeled using approaches like hidden Markov models \citep{cardenas2016markov, gelati2008hidden,hernandez_enso:2020}. Related, \cite{bonner2014modeling} used functional data analysis to study climate teleconnections on land, but did not consider spatial dependencies or connections between locations. Generally, these studies are informative about temporal characteristics but poorly suited to the problem of identifying dependencies.

Alternatively, Complex Network Models have been used to study dependencies in the climate system, leading to the development of the concept of a Climate Network (CN) \citep{tsonis2004architecture, fan2021statistical, ludescher2021network, marwan_networks:2015}. In a CN, the nodes symbolize time series on a spatial grid, while the edges represent the degree of similarity or causality between different grid points. The edges are generated by thresholding pairwise measurements such as Pearson correlation, event synchronization \citep{agarwal2022complex}, and mutual information \citep{donges2009complex}. These approaches, however, may identify spurious edges. For example, a teleconnection might be ``significant"  because two locations are correlated through a chain of local connections; yet this spurious teleconnection should not be included in the later step of network topology analysis. What's more, most of these methods are not suitable for climate indexes with strong auto-correlations \citep{guez2014influence}. Our proposed method aims to address these shortcomings by identifying grid locations that directly influence each other, and modeling time series using smooth functions with substantial flexibility.

The rest of the paper is organized as follows. In Section \ref{sec::def_fgm}, we introduce the proposed dynamic Bayesian functional graphical model. In Section \ref{sec::fgm_simulation}, we compare our model to frequentist and Bayesian competitors using simulated data.  In Section \ref{sec::fgm_application}, we apply the proposed methods to the sea surface temperature data. We conclude in Section~\ref{sec:fgm_conclusions}. The detailed MCMC algorithms are described in the Appendix. All codes to reproduce the results will be available online.

\section{Dynamic Bayesian functional graphical models} 
\label{sec::def_fgm}

\subsection{Bayesian functional graphical models}
\label{bfgm}
The goal of functional graphical models (FGMs) is to learn and express the conditional dependencies between random functions. Unlike traditional graphical models, where the nodes usually represent scalar random variables, the FGM nodes represent random functions on their full domain. $\bm Y^i = \{Y^{i}_{j} (t)\}_{j = 1}^p$ denote iid replicates for $i=1,\ldots,n$ of $p$ random functions $\{y_j\}_{j=1}^{p}$ on the functional domain $t \in \mathcal{T} \subseteq \mathbb{R}$. Since these functions may be observed subject to measurement error, we similarly define $\{X_j^i(t)\}_{j=1}^p$ to be iid realizations of the error-free functions $\{x_j\}_{j=1}^p$. The conditional dependencies among the $p$ random functions are represented by an undirected graph $\mathcal{G}(V, E)$ with vertices  $\mathcal{V} = \{1,\ldots,p\}$ and undirected edges $\mathcal{E} \subset \mathcal{V} \times \mathcal{V}$. An edge $(j_1,j_2)$ exist if and only if the corresponding two functions are \emph{not} conditionally independent. Specifically, we define conditional independence for functional data as in \cite{zhu2016bayesian}: letting $A,B,C \subset \{1,\ldots,p\}$ we say that $\bm x_A$ and $\bm x_B$ are \emph{conditionally independent} given $\bm x_C$ if and only if the conditional distribution $[\bm x_A\mid \bm x_B, \bm x_C]$ equals the conditional distribution $ [\bm x_A\mid \bm x_C]$. 

To represent the functions, we expand each $x_{j}(t)$ on a set of basis functions $\{f_{j,k}(t)\}_{k =1}^{\infty}$. The basis function system is predetermined according to the characteristics of the observed functions, and can take various forms such as functional principal components, B-splines, Fourier basis, or wavelets. We assume that the random functions can be sufficiently represented with truncated basis expansion of level $K_j$. For notation simplicity, we use the same basis $\{f_{k}\}_{k =1}^{\infty}$ and truncation level $K$ for all $j=1,\ldots,p$. The observed random functions are modeled as 

\begin{align}
    \label{truncated-model}
    Y^{i}_{j}(t) &= X_j^i(t) + \sigma_\epsilon\epsilon^{i}_{j}(t), \quad \epsilon^{i}_{j}(t) \stackrel{iid}{\sim} N(0, 1),\\
    \label{error-free}
    X_j^i(t) &= \sum_{k=1}^{K} c^{i}_{j,k} f_{k}(t)
\end{align}
where 
$\{\bm c^{i}_{j}\}_{i=1}^n$ are iid replicates of the basis coefficients  $\bm c_{j} = (c_{j,1} ,\ldots , c_{j,K})'$ for each $x_j$. Through the truncation of basis expansion in \eqref{truncated-model}, we move from an infinite-dimensional space to a finite-dimensional space, and study the conditional dependencies of random functions $\{x_j\}_{j=1}^p$ instead of $\{y_j\}_{j=1}^p$. This strategy is widely used in FGMs and uses $\epsilon^{i}_{j}(t) = Y_j^i(t) - X_j^i(t)$ to represent both measurement errors and truncation errors. 

Let $\bm c = (\bm c_{1}',  \ldots, \bm c_{p}')'$ denote the sequence of all basis coefficients for all $p$ functions. We assume a Gaussian prior on the coefficient vectors:
\begin{equation}
    \bm c \sim MVN(\bm 0, \bm \Omega^{-1}), 
    \label{normal-prior}
\end{equation}
where $\bm \Omega$ is the $(pK)\times (pK)$ precision matrix for the basis coefficients. 

A critical feature of 
model \eqref{truncated-model} and \eqref{normal-prior} is that conditional independence in the basis space corresponds to conditional independence in the functional space:

\begin{equation}
    \bm c_{j_1} \perp \!\!\! \perp 
    \bm c_{j_2} \mid  \bm c_{j}, \forall j \neq j_1, j_2 
    \iff  x_{j_1} \perp \!\!\! \perp x_{j_2} \mid  x_{j},  \forall j \neq j_1, j_2 
    \iff (j_1, j_2) \not \in E.
    \label{coeff-space-indep}
\end{equation}

The fundamental idea of \emph{Gaussian} graphical models is that 
conditional independence corresponds to sparsity in the precision matrix. 
Let $\bm \Omega_{j_1, j_2} = \{\omega_{j_1 k_1, j_2 k_2}\}_{k_1, k_2 = 1}^K$ denote the $(j_1, j_2)$th block of $\bm \Omega$ for  $j_1, j_2 = 1, \ldots, p$, so each block is $K \times K$. Under the Gaussianity assumption \eqref{normal-prior}, it is easy to show that a zero block $\bm \Omega_{j_1, j_2} = \bm 0_{K \times K}$ is equivalent to the conditional independence between $\bm c_{j_1}$ and $\bm c_{j_2}$ in \eqref{coeff-space-indep}. Therefore, we deduce that,  based on \eqref{coeff-space-indep}, a zero block $\bm \Omega_{j_1, j_2} = \bm 0_{K \times K}$ is equivalent to
the absence of an edge between nodes $j_1$ and $j_2$ in 
the functional space $\mathcal{G}(V, E)$. In other words, learning the edge set $E$ of the graph $\mathcal{G}(V,E)$ is equivalent to learning the nonzero blocks of $\bm \Omega$. In order to recover edges of the graph $\mathcal{G}$ in the functional space, we must learn the \emph{block-sparsity} structure of $\bm \Omega$. 

In Bayesian graphical models, the sparsity of the precision matrix is encoded in the prior distribution of $\bm \Omega$. We build upon the continuous spike-and-slab model of \cite{wang2015scaling} and propose a novel prior with a block-wise sparsity structure. Define a basis space graph $\mathcal{G}^c$ for $\bm c$ via the edge inclusion indicators $\bm G = \{g_{j_1k_1, j_2 k_2}\}$ for $g_{j_1k_1,j_2k_2} \in \{0,1\}$. These indicators will be modeled such that the precision entries are (near) zero, $\omega_{j_1k_1, j_2k_2} \approx 0$, whenever $g_{j_1k_1,j_2k_2} = 0$. Thus, block-wise sparsity will require joint consideration of many indicators $\{g_{j_1k_1,j_2k_2}\}_{k_1,k_2=1}^K$: if all indicators for the $(j_1, j_2)$ block are simultaneously zero, then we infer conditional independence between the nodes (i.e., functions) $j_1$ and $j_2$.

First, conditional on $\bm G$, the prior distribution for the precision matrix is 
\begin{equation}
    \label{prec-prior}
        p(\bm \Omega \mid  \bm G, v_0, v_1, \lambda)  
    \propto
    \prod_{i_1<i_2} N( \omega_{i_1 i_2} \mid  0, v_{g_{i_1 i_2}}^2 ) \cdot
    \prod_i \textrm{Exp}( \omega_{ii} \mid  \lambda/2) \cdot
    \bm 1_{\bm \Omega \in M^{+}},
\end{equation}
where $i_1$ and $i_2$ indicate all the off-diagonal elements in $\bm \Omega$, $\textrm{Exp}(\omega_{ii}\mid  \lambda/2)$ denotes the exponential distribution with rate $\lambda/2$, $\bm 1$ is the indicator function, and $M^+$ is the space of positive definite matrices. The prior of $w_{j_1k_1,j_2k_2 }$ can be considered as a mixture of a normal distribution with a small variance $v_0^2$ (spike) and a normal distribution with a large variance $v_1^2$ (slab), with $v_0 \ll v_1$. The case $g_{j_1k_1, j_2 k_2} = 0$ implies that $\omega_{j_1k_1, j_2 k_2}$ belongs to the spike component with a small variance, thus its value is effectively zero. The case $g_{j_1k_1, j_2 k_2} = 1$ implies that $\omega_{j_1k_1, j_2 k_2}$ belongs to the slab component with a larger variance, thus its value is more substantially nonzero. This prior extends \cite{wang2015scaling} to the functional data setting. The positive definiteness of the precision matrix is guaranteed in our proposed MCMC algorithm by Sylvester’s criterion. More details are given in Appendix \ref{sec::mcmc_dbfgm}.
 
Next, we introduce block-wise edge inclusion probabilities $\bm \pi = (\pi_0, \{\pi_{j_1 j_2}\}_{ 1 \leq j_1<j_2 \leq p} )$, where $\pi_0$ is the edge inclusion probability for the diagonal blocks $\bm \Omega_{j,j}$, and $\pi_{j_1 j_2}$ the edge inclusion probabilities of the off-diagonal blocks $\bm \Omega_{j_1,j_2}$. Specifically, the joint prior on $(\bm G, \bm \pi)$ is  
\begin{align}
\begin{split}\label{graph_prior}
    p(\bm G \mid \bm \pi) 
    & \propto \prod_{j_1<j_2} 
     \prod_{k_1, k_2} 
     \pi_{j_1 j_2}^{g_{j_1 k_1, j_2 k_2}} (1-\pi_{j_1 j_2})^{(1-g_{j_1 k_1, j_2 k_2})}\\
    & \quad \quad \times \prod_{j=1}^p 
     \prod_{k_1<k_2} 
     \pi_{0}^{g_{j k_1, j k_2}} (1-\pi_{0})^{(1-g_{j k_1, j k_2})}
\end{split} \\
    p(\bm \pi) 
    &\propto
    \textrm{Beta}(\pi_0, \mid  \alpha_0, \beta_0)
    \prod_{j_1<j_2} \textrm{Beta}(\pi_{j_1, j_2}\mid  \alpha, \beta).
    \label{pi_prior}
\end{align}
Crucially, this prior incorporates a hierarchical block-wise sparsity structure for $\bm G$. As $\pi_{j_1 j_2}$ tends toward zero, the edges in the $(j_1, j_2)$th block of $\bm G$ are increasingly likely to be all zeros. This corresponds to a lack of an edge $(j_1, j_2)$  in the functional space graph $\mathcal{G}$. Unlike previous approaches that use a single weight for the spike-and-slab mixture components across all edges \citep{wang2015scaling},  our mixture weights $\pi_{j_1 j_2}$ are (i) block-specific and (ii) assigned hierarchical Beta priors, thus learned from the data. This prior \eqref{graph_prior}--\eqref{pi_prior} is carefully designed for data-adaptive, block-wise sparsity in $\bm G$, which leads to sparse and accurate estimation of the function space graph $\mathcal{G}$ (see Section \ref{sec::fgm_simulation}).

Section \ref{sec::choice-of-parameter} provides supplementary analysis of the prior based on Monte Carlo samples.  The hyperparameters $\alpha$ and $\beta$ in the Beta distribution control the level of sparsity in the graph $\bm G$, which is then used to infer the graph $\mathcal{G}$. When the mean is higher, the resulting graphs in the basis space (Figure~\ref{prior-inclusion-probability}) and the functional space (Figure~\ref{prior-inclusion-probability-alt}) are denser. In the simulation study and the real data analysis, we assume that the on-diagonal blocks are always fully connected and fix $\pi_0 = 1$ for simplicity;  results are not sensitive to this choice. Lastly, we specify an inverse-Gamma prior for $\sigma_{\epsilon}^2$ with shape $\alpha_{\sigma}$ and rate $\beta_{\sigma}$.

The noteworthy features of model \eqref{truncated-model}--\eqref{pi_prior} are 
(i) the joint prior encourages block-wise sparsity of the precision matrix, thus enabling graph estimation in the functional space; 
(ii) the prior on the basis coefficients $\bm c$ introduces shrinkage and regularization via elementwise sparsity of $\bm\Omega$, which encourages both parsimony and information-sharing among the basis coefficients to improve estimation and inference in the functional space; 
and (iii) the basis expansion and graph prior are amenable to efficient computing via a Gibbs sampling algorithm (see Section~\ref{sec-mcmc}). As such, this model represents a compelling alternative to existing Bayesian FGMs \citep{zhu2016bayesian}. In addition, the model proposed in \cite{zhu2016bayesian} is restricted to decomposable graphs while our model does not have this constraint.

\subsection{Dynamic Bayesian functional graphical models}

The static Bayesian functional graphical model from Section \ref{bfgm} can be applied to functions defined over any compact continuum such as time, space, and wavelength. In this section, we consider functions over a time interval, and introduce dynamics of the graph within this time interval. To avoid confusion, note that the observed data are iid realizations of multivariate random functions $\bm y = \{y_j (t)\}_{j = 1}^p$, where $t$ represents time. This setting is different from a time sequence of random functions, where a whole function  (or a set of functions) over some continuum is observed at every time point. However, the main ideas apply for modeling changes over any one-dimensional compact domain $\mathcal{T}$ on which the functions are defined, such as space or wavelength.

We introduce graph dynamics via an unknown \emph{changepoint}  $\tau \in \mathcal{T}$, with different graphs before and after the changepoint. The model is specified to allow the basis coefficients and the error variances to differ before and after $\tau$:

\begin{equation}\label{truncated-model-cp}
\begin{split}
    Y^{i}_{j}(t) &= 
    \sum_{k=1}^{K} c^{i(1)}_{j,k} f_{k}(t) + \sigma_\epsilon^{(1)}\epsilon^{i}_{j}(t), \quad t<\tau,\\
     Y^{i}_{j}(t) &= 
    \sum_{k=1}^{K} c^{i(2)}_{j,k} f_{k}(t)+ \sigma_\epsilon^{(2)}\epsilon^{i}_{j}(t), \quad t \geq \tau
\end{split}
\end{equation}
with $\epsilon^{i}_{j}(t)$ distributed as in \eqref{truncated-model} and the error-free functions 
$X_j^i(t) = \sum_{k=1}^{K} c^{i(s)}_{j,k} f_{k}(t)$ defined as in \eqref{error-free}, now with $s = 1$ for $t < \tau$ and $s = 2$ for $t \ge \tau$. 
Similar to the static version of the FGM in \eqref{normal-prior}, 
$\bm c^{i(1)}_{j}$ and $\bm c^{i(2)}_{j}$ are iid realizations of $\bm c^{(1)}_{j}$ and $\bm c^{(2)}_{j}$, respectively, and the concatenated representation of basis coefficients before and after the changepoint are 
$\bm c^{(1)} = (\bm c^{(1)'}_{1}, \ldots, \bm c^{(1)'}_{p})'$, $\bm c^{(2)} = (\bm c^{(2)'}_{1}, \ldots, \bm c^{(2)'}_{p})'$. The basis coefficients have multivariate Gaussian priors with different precision matrices before and after the changepoints:
\begin{equation}
     (\bm c^{(1)}, \bm c^{(2)}
    ) \sim
    MVN(\bm c^{(1)} \mid  \bm 0, [\bm \Omega^{(1)}]^{-1})\cdot
     MVN(\bm c^{(2)} \mid  \bm 0, [\bm \Omega^{(2)}]^{-1}).
         \label{normal-prior-cp}
 \end{equation}

The crucial feature of model \eqref{truncated-model-cp}--\eqref{normal-prior-cp} is that it defines a \emph{dynamic} graph via the sparsity patterns in $\bm \Omega^{(1)}$ and $\bm \Omega^{(2)}$ and the location (in time) of the changepoint $\tau$, which partitions the functional domain $\mathcal{T}$. By revisiting the correspondence \eqref{coeff-space-indep}, we see that $\bm \Omega^{(1)}$ and $\bm \Omega^{(2)}$ define two edge sets, $E^{(1)}$ and $E^{(2)}$, based on the conditional dependencies among the nodes. These nodes correspond to the same $p$ random functions, but now \emph{restricted to each partition}, $\{x_j(t):t < \tau\}_{j=1}^p$ and $\{x_j(t):t \ge \tau\}_{j=1}^p$, respectively. Thus, within each partition, the traditional FGM graph definition applies: the nodes are the (restricted) random functions and edges correspond to conditional dependencies between nodes. Notably, these distinct graphs are self-contained: the conditional dependencies among $\{x_j(t):t < \tau\}_{j=1}^p$ do \emph{not} require conditioning on $\{x_j(t):t \ge \tau\}_{j=1}^p$ (or vice versa). Each precision matrix $\bm \Omega^{(1)}, \bm \Omega^{(2)}$ is assigned the same prior \eqref{prec-prior}  with accompanying edge inclusion indicators $\bm G^{(1)}, \bm G^{(2)}$ and block-wise edge inclusion probabilities $\bm \pi^{(1)}, \bm \pi^{(2)}$, also using the same priors \eqref{graph_prior}--\eqref{pi_prior} as before. The error variances $\sigma_\epsilon^{(s)2}$, $s = 1,2,$ before and after the changepoint follow independent inverse-Gamma priors with shape $\alpha_\sigma$ and rate $\beta_\sigma$.

Finally, we specify a uniform prior for the changepoint, $\tau \sim \textrm{Uniform}( \tau_{min}, 
    \ldots, \tau_{max} )$, 
which assigns equal point mass to every time point between a specified time interval  $[\tau_{min}, \tau_{max}]$. Here,
$\tau_{min} > 1$ and $\tau_{max} < T$ are the minimum and maximum boundaries, respectively, and can be specified in applications based on domain knowledge. This prior allows us to estimate the changepoint via the joint posterior distribution. 

These modeling strategies are directly generalizable to multiple (unknown) changepoints. Instead of having one changepoint and two graphs $\bm G^{(1)}$ and $\bm G^{(2)}$, the model can be specified to have $S-1$  changepoints $(\tau^{(1)}, \tau^{(2)}, \ldots, \tau^{(S-1)})$ and $S$ graphs $\{\bm G^{(s)}\}_{s=1}^{S}$. The changepoints  are defined on the space of all possible ordered tuples, $1<\tau^{(1)} < \tau^{(2)} < \ldots < \tau^{(S-1)}<T$. The prior is a simple uniform prior on all possible tuples. A user can specify constraints on each changepoint based on domain knowledge. Appendix \ref{sec::mcmc_dbfgm} contains the full mathematical formulation of this model.

\subsubsection{MCMC algorithm}\label{sec-mcmc}

We design an efficient Markov chain Monte Carlo (MCMC) algorithm to sample the model parameters from their joint posterior distribution. In the dynamic Bayesian FGM, the unknown parameters are $\tau$ and $\{\bm G^{(s)},\bm \Omega^{(s)},\bm \pi^{(s)},\bm c^{i(s)}, \sigma_\epsilon^{(s)}\}$ for $s=1, 2$. Importantly, the graph dynamics induced by the changepoint maintain many of the same sampling steps as in the static Bayesian FGM  from Section~\ref{bfgm}, and extensions for multiple changepoints are straightforward. 
A generic iteration of the Gibbs sampler uses the following steps for each $s = 1, 2$, with details provided in Appendix \ref{sec::mcmc_dbfgm}:
 \begin{itemize}   
    \item[]  \textbf{a. Sample $\bm \Omega^{(s)}$:}
    Conditioned on the graph $\bm G^{(s)}$ in the coefficient space, we update the precision matrix ${\bm \Omega^{(s)}}$ using a block Gibbs sampler with closed-form conditional distributions for each column. The sampler automatically guarantees positive definiteness of the precision matrix, see the Appendix. 
    
    \item[]  \textbf{b. Sample $\bm G^{(s)}$:}
    We update $\bm G^{(s)}$ drawing from independent Bernoulli distributions, where the probability of each edge inclusion is determined by the block-wise edge inclusion probability $\bm \pi^{(s)}$ .
    
    \item[] \textbf{c. Sample $\bm \pi^{(s)}$:} Given graph $\bm G^{(s)}$, the posterior of each element in $\bm \pi^{(s)}$ is a conjugate beta distribution. 

    \item[] \textbf{d. Sample $\{\bm c^{i(s)}\}$, $i = 1, \ldots, n$ and $\{\sigma_\epsilon^{(s)}\}$:} For each $i$, sample the basis coefficients in each state from a conjugate multivariate normal distribution. Then sample $\{\sigma_\epsilon^{(s)}\}$ from an inverse-Gamma distribution.

    \item[] \textbf{e. Sample $\tau^{(s)}$}: Since the prior is a uniform distribution on the observed time points $\{\tau_{min},\ldots,\tau_{max}\}$, the posterior distribution is a discrete distribution on $\{\tau_{min},\ldots,\tau_{max}\}$. 
\end{itemize}


Our proposed model requires a fixed level of truncation $K$ and a fixed number of changepoints. In the application of this paper, these two hyperparameters are selected using  deviance information criterion (DIC). The model remains applicable with distinct truncation levels $K_j$ for each function $j=1,\ldots,p$, but then requires a choice of each $K_j$. For simplicity, we prefer a single choice of $K_j = K$, and inspect the fitted curves to ensure that the chosen basis is sufficiently flexible to capture the dominant trends in the observed functions.

\subsubsection{Posterior inference}
\label{sec::posterior-inference-method}
The primary inferential target is the dynamic graph in the functional space. Specifically, we leverage the joint posterior distribution from the proposed model to estimate the edge sets $E^{(1)}$ and $E^{(2)}$ before and after the changepoint, respectively, which describe the conditional dependencies among the random functions on the respective subdomains $\{t < \tau\}$ and $\{t \ge \tau\}$. For each $s=1,2$, we first compute the marginal posterior edge inclusion probabilities of $\bm G^{(s)}$, which represents the $(pK)\times (pK)$ graph in the coefficient space, using the posterior samples from Section~\ref{sec-mcmc}. More precisely, the posterior probability $p(g^{(s)}_{j_1 k_1,j_2 k_2}=1\mid Y^1, \ldots, Y^n)$ is estimated as the proportion of MCMC samples with $g^{(s)}_{j_1 k_1,j_2 k_2}=1$. Then the \emph{posterior median graph} of $\bm G^{(s)}$  is estimated by thresholding these probabilities by 0.5. Finally, we convert this graph estimate to the functional space via the block-wise correspondence from \eqref{coeff-space-indep}: an edge $(j_1, j_2)$ exists if and only if the corresponding $(j_1, j_2)$th block in $\bm G^{(s)}$ is not a zero block, for $j_1,j_2=1,\ldots,p$. 

An alternative approach for functional graph estimation is to convert each posterior sample of $\bm G^{(s)}$ into a function space graph $\mathcal{G}^{(s)}$, and then compute the posterior median of $\mathcal{G}^{(s)}$ directly. However, the posterior edge inclusion probabilities under this approach are typically much larger than for the previous approach, and often much larger than 0.5.  Thus, higher threshold probabilities are required to obtain sparse functional graph estimates. We prefer the previous approach because it more directly leverages the hierarchical block-wise sparsity prior for $\bm G$ and produces sparse yet accurate graph estimates for $\mathcal{G}$.

\subsubsection{Choice of hyperparameters}
\label{sec::choice-of-parameter}
Bayesian graphical models commonly set the prior marginal edge inclusion probability to $2/(p-1)$, which implies that the prior expected number of edges is $p$. Accordingly, we set the prior belief of edge inclusion probabilities of graph $\bm G^{(s)}$ in the coefficient space to $2/(p-1)$, corresponding to an expected number of edges equal to $pK^2$. As in \cite{wang2015scaling} and \cite{peterson2020bayesian}, the prior distribution of graph $\bm G$ is not analytically available because of  intractable normalizing constants. A complete form of the full prior with the normalizing constants is included in Appendix \ref{sec::mcmc_dbfgm}. In order to investigate whether the normalizing constants in our new model dominate the prior and drive the prior to the extreme, we evaluate the prior edge inclusion probability based on Monte Carlo sampling from the prior.

\begin{figure}
\centering
    \includegraphics[width=0.4\textwidth]{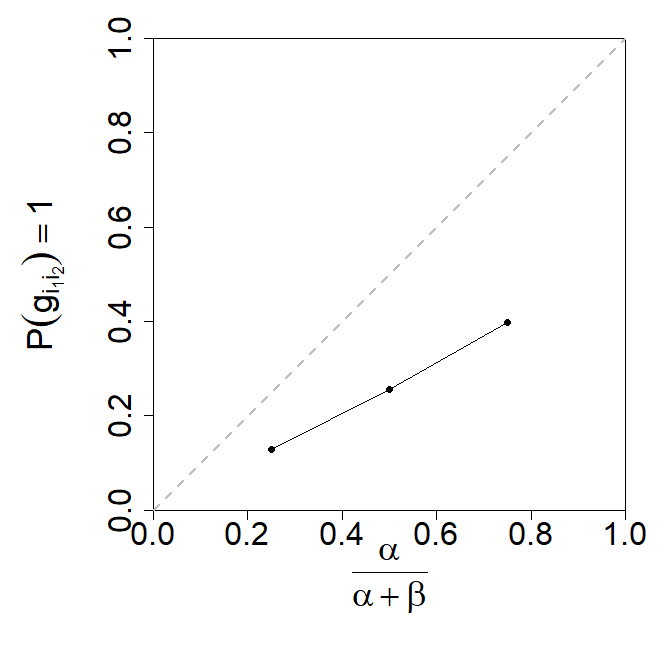}
    \includegraphics[width=0.4\textwidth]{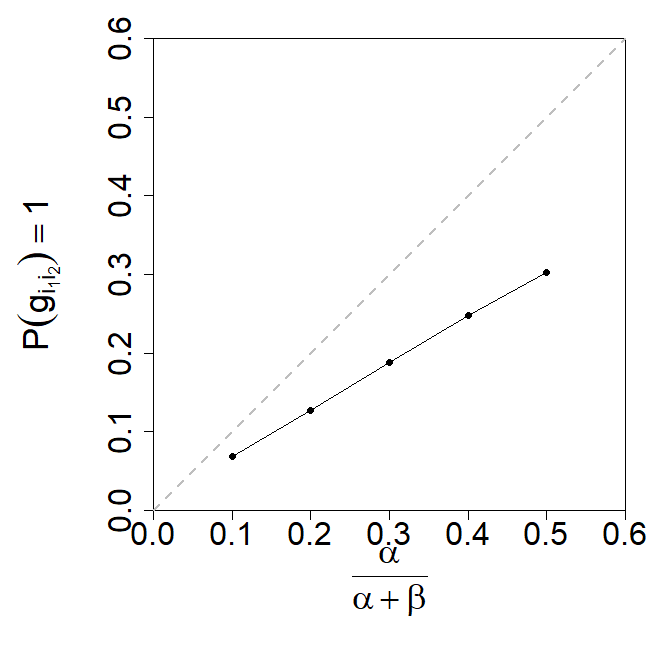} \\
    \includegraphics[width=0.4\textwidth]{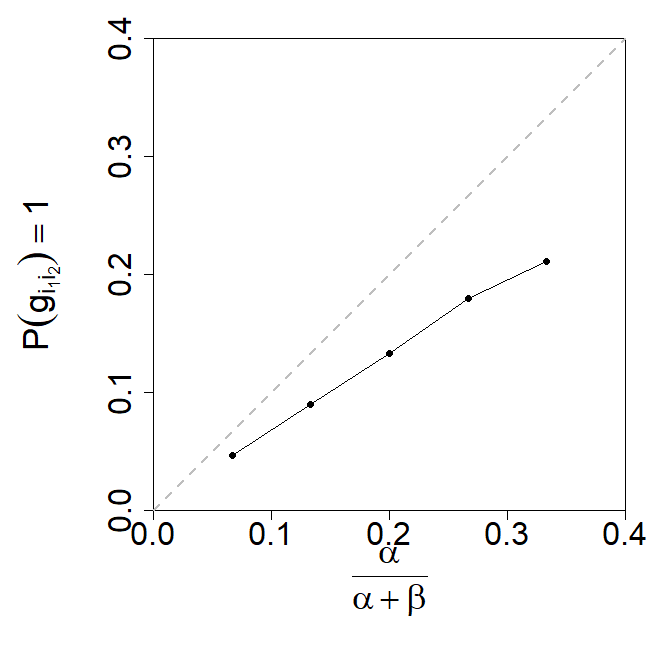} 
    \includegraphics[width=0.4\textwidth]{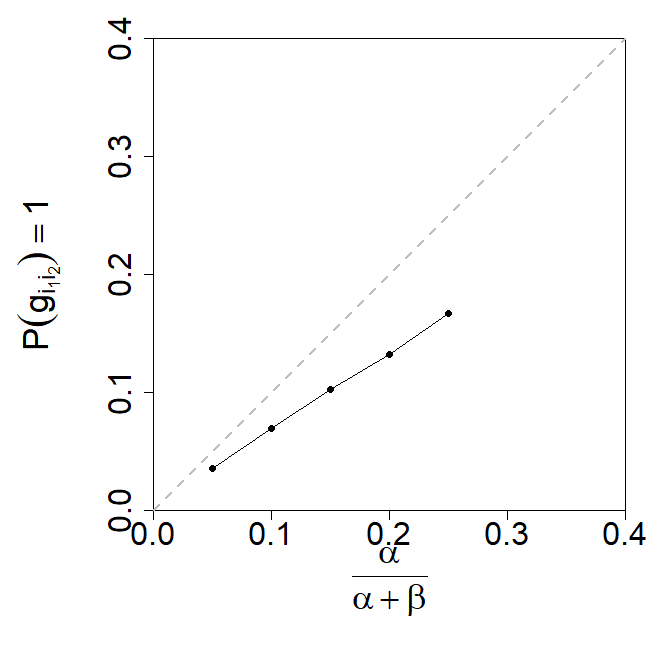}
        \caption{Prior edge inclusion probability (y-axis) versus mean of the Beta prior, $\alpha/(\alpha + \beta)$, (x-axis) for $p=5$ (top left), $p=10$ (top right), $p=15$ (bottom left), and $p=20$ (bottom right). The dotted gray line is when the two axes are equal.}
    \label{prior-inclusion-probability}
\end{figure}

First, we evaluate the influence of $\alpha$ and $\beta$ on the marginal prior edge inclusion probability. We fix $\lambda = 1$, $\nu_0 = 0.02$ and $h=50$ ($v_1 = h  v_0$) as commonly recommended \citep{wang2015scaling,peterson2020bayesian,osborne2022latent}. We also fix $\alpha = 2$ and $K=5$. The value of $\beta$ varies so that the prior mean of the beta distribution kernel for each $\pi_{j_1, j_2}$ in \eqref{pi_prior} 
takes values in $\{1/(p-1), 2/(p-1), 3/(p-1), 4/(p-1), 5/(p-1)\}$. The study is carried out on $p = (5, 10, 15, 20)$. Figure~\ref{prior-inclusion-probability} shows the prior edge inclusion probabilities estimated via Monte Carlo sampling. We conclude that: 1) the prior edge inclusion probability is not dominated by the normalizing constant; 2) similar to \cite{wang2015scaling}, the prior edge inclusion probability is lower than the prior mean of $\pi_{j_1, j_2}$; and 3) the influence of normalizing constants on the prior edge inclusion probability is greater when $p$ is smaller, but the differences among different values of $p$ are negligible. Figure~\ref{prior-inclusion-probability} can be used as a guide to choose parameters that reflect the preferred edge inclusion probability. 
In the example of $p=15$ the prior edge inclusion probability is slightly below $2/(p-1)$ when the prior mean of $\pi_{j_1,j_2}$ is $3/(p-1)$. Therefore, when $p = 15$, $\alpha = 2$ and $\beta = 7$ is a reasonable choice.

Next, we fix $\alpha = 2$, $\beta=7$, $v_0 = 2$ and $p = 15$, and compute the prior edge inclusion probability given $K = (3, 5, 10)$ and $h = (10, 50, 100)$. The values of edge inclusion probabilities under each circumstance are shown in Appendix \ref{sec::prior-evaluation} (Table \ref{tab:prior-inclusion-table}) and show that the probability decreases as $K$ and $h$ increase, which is consistent with the findings in \cite{wang2015scaling}. Lastly, we examine if the choice of $\nu_0 = 0.02$ and $h = 50$ results in a clear separation between the spike component and the slab component, using the values recommended in \cite{wang2015scaling}. The prior distributions of elements of the precision matrix, under the spike and slab components, are plotted in Appendix \ref{sec::prior-evaluation} (Figure \ref{precision-prior}) and show separation between the two groups, with group zero small enough to be estimated as zero.  

\section{Simulation study}
\label{sec::fgm_simulation}

\subsection{Data generation} \label{data_generation}
\begin{figure}
    \centering
    \includegraphics[width=1.5in,height=1.5in]{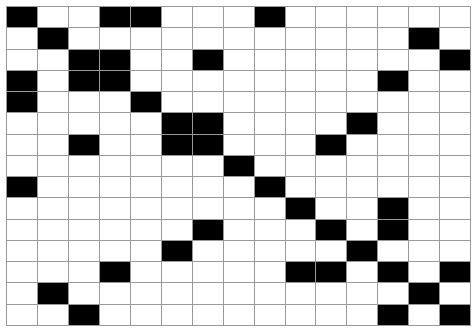}
    \includegraphics[width=1.5in,height=1.5in]{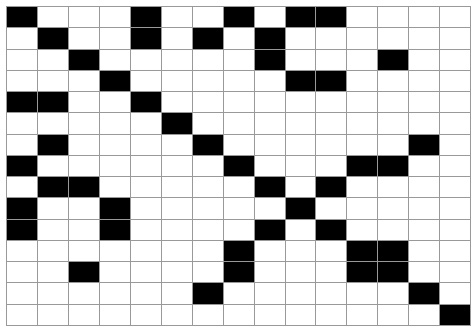}
    \caption{
    Simulated $p \times p$ graph in the functional space before (left) and after (right) the changepoint. Black block indicates that the two functions are connected by an edge.}
    \label{simulated data}
\end{figure}

To assess the performance of our approach relative to competing methods, we generate synthetic multivariate functional data from a dynamic functional graphical model. Specifically, we generate $n = 50$ replicates of $p=15$ random functions on an equally-spaced grid of $T=256$ observation points. The functions are generated from the changepoint basis expansions \eqref{truncated-model-cp} with true changepoint $\tau = 129$, polynomial basis functions of order 4 ($K=5$), and error standard deviation 0.05. To simulate the basis coefficients---and incorporate a dynamic graph among the functions---we first generate graphs pre- and post-changepoint in the functional space.  Both $p\times p$ graphs are simulated with independent edge inclusion probabilities $2/(p-1)$ and presented in Figure~\ref{simulated data}. The block-wise sparsity of each basis coefficient graph is induced by the sparsity of the corresponding functional graph. The nonzero elements of the basis coefficient precision matrices are then simulated from a G-Wishart distribution with degrees of freedom set to 3 and scale matrix set to the identity matrix, and the basis coefficients are finally simulated from \eqref{normal-prior-cp}. Prior to adding the iid errors in \eqref{truncated-model-cp}, we apply a continuity correction at $\tau$ so that $\sum_{k=1}^K c^{i(1)}_{j,k} f_{k}(\tau) = \sum_{k=1}^K c^{i(2)}_{j,k} f_{k}(\tau)$. Specifically,  a constant is added or subtracted from each segment so that the values of two adjacent segments are equal at the changepoint.

Unlike the common setup in existing graphical model or functional graphical model studies, our precision matrices do not have a predefined structure, such as a auto-regressive structure, a  block-diagonal structure, or a special design that ensures the off-diagonal elements  have large absolute values. Instead, our graphs in the functional space are dynamic and unstructured, which produces a rich and challenging collection of simulated datasets. 


\subsection{Competing methods}
To compare with the proposed dynamic Bayesian functional graphical model (DB-FGM), we include a variety of state-of-the-art graphical models---including Bayesian and non-Bayesian, dynamic and static, and functional and traditional graphical models. First, we include the partially separable functional graphical model (PS-FGM; \citealp{zapata2019partial}), which is a frequentist FGM. Since the PS-FGM model estimates one static graph, we supply the information on the \emph{true} changepoint and estimate two graphs separately before and after the changepoint. Naturally, this knowledge of $\tau$ provides a significant advantage. Otherwise, we use the default settings from the \texttt{fgm} package in \texttt{R}.

Second, we include  the Bayesian functional graphical lasso \citep{zhang2021bayesian}, which is a recent dynamic and Bayesian FGM. We use the same B-spline basis as in the DB-FGM (see Section~\ref{sec:param}) and the vague prior from \cite{zhang2021bayesian}. Although this model produces dynamic graphs, the pairwise dependencies are always conditional on functions over the whole domain. Thus, we consider two variations: (i) direct application to the observed functional data (BL-FGM-full) and (ii) separate application to data pre- and post-changepoint  (BL-FGM-indp). Again, knowledge of $\tau$ provides an advantage for  BL-FGM-indp. 

As a final competitor, we fit the Bayesian Gaussian graphical model (B-GGM) proposed by \cite{wang2015scaling} independently at each time point $t \in \mathcal{T}$. The resulting graphs are dynamic, but there is no information-sharing across the functional domain $\mathcal{T}$. It is worth mentioning that the dynamic graphical models with changepoint detection of \cite{liu2022dynamic} and \cite{franzolini2024change} are not suitable for comparison to our model
as they treat the data as a single time series instead of replications, as we do, and cannot guarantee that the
same (or even similar) graphs and changepoints will repeat over the years.

\subsection{Parameter settings} \label{sec:param}

We use $K=5$ B-splines of order 3 to fit the model, which provides accurate estimation of the underlying true smooth curves. Note that this basis differs from the polynomial basis used to generate the data. 
For the hyperparameters in the joint prior \eqref{prec-prior}, we follow the guidelines in Section \ref{sec::choice-of-parameter} and set $\lambda = 1$, $v_0 = 0.02$, $h = 50$ ($v_1 = h  v_0$), $\alpha = 2$ and $\beta=7$. To initialize the MCMC, the precision matrix and graphs are all set to the identity matrix and the block-wise edge inclusion probabilities are set to 0.5. Our simulation study shows that the model is robust to the initialization of the basis coefficients and the changepoint $\tau$. Results reported below were obtained by running MCMC chains for 5000 total iterations with a burn-in of 3000 iterations. The models took on average about 4 minutes on a Mac computer with a 16 GB Apple M1 Pro chip. MCMC convergence was checked by the Geweke diagnostic test \citep{geweke1991evaluating} on the key parameters $\{\bm c^{i(s)}, \sigma_\epsilon^{(s)}, \tau\}$.

\subsection{Results}
\label{sec::simulation-results}

For each graph before and after the changepoint, we assess performance on edge selection using the true positive rate (TPR), the false positive rate (FPR) and the Matthews correlation coefficient (MCC). The MCC is defined as 
\begin{equation*}
    MCC = \frac{TP \times TN - FP \times FN}{
    \sqrt{(TP + FP) \times (TP + FN) \times (TN + FP) \times (TN + FN)}}
\end{equation*}
where TP is the number of true positives, TN is the number of true negatives, FP is the number of false positives, and FN is the number of false negatives. The MCC provides a measure of overall classification success for a selected model, with larger values indicating better performance. Since the Bayesian functional graphical lasso models (BL-FGM-indp and BL-FGM-full) and the B-GGM estimate a graph at each time point $t$, we compute the averaged rates across all time points as suggested by \cite{zapata2019partial}. In other words, the estimated graph at each time $t$ is evaluated against the true graph $G^{(s_t)}$ at time $t$; the metrics are then averaged over all times $t$  within each segment. This approach is commonly used to evaluate dynamic graphical models over a time period. Notably, the estimated time-varying graphs are \emph{not} summarized into a single, static graph for each segment. The results averaged across 50 simulated datasets are presented in Table~\ref{table-performance-fgm}.The comparative performances are very consistent across replications.The results of BL-FGM-indp and BL-FGM-full were computed from 25 simulations, as these methods are computationally very intensive and their performance rates had small variances across replications.

\begin{table}[H]
{
\centering
\begin{tabular}{|c c c c c c |}  
\hline
 &  DB-FGM &
PS-FGM& 
BL-FGM-indp &
BL-FGM-full&
B-GGM\\ 
\hline
\multicolumn{1}{|c}{} &  \multicolumn{5}{c|}{$s=1$}\\ 
 \hline
 TPR & {\bf 0.72} (0.1) & 0.42 (0.12)  & 0.44 (0.01) & 0.51 (0.01) & 0.57 (0.02)\\
FPR & {\bf 0.04} (0.02) & 0.07 (0.04)  & 0.09 (0.004) & 0.10 (0.002)& 0.08 (0.01)\\
MCC & {\bf 0.68} (0.1) & 0.38 (0.11)   & 0.35 (0.01) & 0.39 (0.001) & 0.48 (0.09)\\\hline
\multicolumn{1}{|c}{} &  \multicolumn{5}{c|}{$s=2$}\\
 \hline
TPR & {\bf 0.87} (0.1) & 0.24 (0.12)  & 0.19 (0.01) & 0.06 (0.01) & 0.12 (0.01)\\
FPR & 0.07 (0.02) & {\bf 0.03} (0.02)   & 0.10 (0.01) & 0.05 (0.01)& 0.09 (0.01)\\
MCC & {\bf 0.73} (0.07) & 0.3 (0.12)  & 0.11 (0.01)  & 0.03 (0.01)& 0.05 (0.02)\\
\hline
\end{tabular}               
\caption{True positive rate (TPR), false positive rate (FPR), and Matthews correlation coefficient (MCC) for graph estimation before ($s=1$) and after ($s=2$) the changepoint. The proposed approach (DB-FGM) significantly outperforms the competitors by nearly all metrics. 
}  
\label{table-performance-fgm}
}
\end{table} 

Results in Table~\ref{table-performance-fgm} show that the proposed DB-FGM model outperforms all competitors across nearly all metrics, often by a wide margin. DB-FGM is significantly more powerful to detect true edges (large TPR), yet incurs very few false positives (small FPR). The estimates of the post-changepoint graph from competing methods are sparser, which suggests that these connections are weaker than those pre-changepoint and more difficult to identify. While all Bayesian FGMs (BL-FGM-indp, BL-FGM-full and B-GGM) generate nearly empty graphs post-changepoint, DB-FGM is able to detect these weaker connections.  Further, DB-FGM achieves this excellent performance without knowledge of the true changepoint, and in fact learns this parameter accurately (posterior mean 129, standard deviation 0.8). By comparison, both PS-FGM and BL-FGM-indp use the true changepoint, which is not available in real data analysis. 

For further comparison, we also fit the static version of the proposed Bayesian FGM from Section~\ref{bfgm}.  The estimated (static) graph is very dense with a large FPR (not shown). This result is unsurprising: because  FGMs consider random functions over the \emph{whole} domain, nodes that are connected only pre-changepoint but not post-changepoint (or vice versa) should be included in the static FGM edge set. 
The changepoint-based DB-FGM is highly capable of learning \emph{sparse} and \emph{dynamic} graphs among functional data, and in doing so provides simpler and more parsimonious explanations. 

This simulation study is also informative about how different models perform when the ground truth is the static scenario without any changepoints. Each segment can be considered as a static scenario and the performance of each model in that segment from Table \ref{table-performance-fgm}  can be used for model comparison. The reason is that PS-FGM, BL-FGM-indp and B-GGM all infer the graphs independently within each segment with given changepoints. Since the changepoint of the proposed DB-FGM converges quickly to the truth, its simulation result in each segment can also be considered from a static perspective. Note also that the static competitor (PS-FGM) is supplied the true changepoint, and thus is an effective static benchmark. For real data analysis, we recommend DIC to choose between the proposed static and dynamic model.

We also investigated simulation settings where the position of the true changepoint changed from the mid-point ($\tau = 129$) to other, more extreme, time points. We observed that the MCMC maintained quick convergence to the true changepoint. 

Lastly, we considered simulated data with two or three changepoints using the multiple changepoint extensions of the proposed approach. Again, the model quickly inferred the true changepoints. In conclusion, our proposed method can accurately estimate both graphs and changepoints.

\subsection{Learning the block-sparsity}

As suggested by our simulation results, our proposed model, together with the choice of hyper-parameters, leads to sparse posterior estimation of the $p$-by-$p$ graphs in the functional space. Our prior on $\bm G$ does not require specific edges  to be included or excluded, but rather gives a common marginal (over all  other parameters) probability of inclusion to all edges. 
These parameters induce group shrinkage and enable information sharing among edges within a block. If the data indicate that most of the edges in a block are excluded, then we expect the posterior of $\pi_{j_1 j_2}$ to concentrate on small values, which will further discourage the inclusion of other edges in that block. 

As further empirical evidence, recall that the posterior median graph is determined by the marginal edge inclusion probabilities of $\bm G$ in the coefficient space. The graph in the functional space is then inferred from this posterior estimation of $\bm G$. The posterior conditional inclusion probability of an edge in $\bm G$ in the MCMC algorithm is 
\begin{equation*}
    p(g_{j_1k_1, j_2k_2}=1\mid \bm \pi, \bm \Omega)=
\dfrac{N(\omega_{j_1k_1,j_2k_2}\mid 0,v_1^2)\pi_{j_1,j_2}}
{N(\omega_{j_1k_1, j_2k_2}\mid 0,v_1^2)\pi_{j_1 j_2}+N(\omega_{j_1k_1,j_2k_2}\mid 0,v_0^2)(1-\pi_{j_1 j_2})},
\end{equation*}
which is a combination of  $\pi_{j_1, j_2}$ and the likelihood. In our simulation study, the posterior mean of $\pi_{j_1, j_2}$ was about 0.08 for the zero blocks and 0.19 for the nonzero blocks.  Edges with a high likelihood in the slab component have a high posterior marginal inclusion probability, which results in an estimated edge in the functional space. To illustrate this point, we calculated the Frobenuius norm of estimated precision matrix $\bm\hat{\bm\Omega}$ in each block, and compared the distribution of these Frobenius norms between zero and nonzero blocks (see Appendix~\ref{sec::prior-evaluation}). There is a clear separation between the two groups. The density from the zero blocks is highly concentrated around zero, especially compared to the density from the nonzero blocks. 




\subsection{Setting $p>n$}
We studied additional settings that included $p>n$ cases, i.e., for $p \in \{20, 30, 40, 50\}$ and $n = 30$, under the same parameter settings as described above. As expected, the graph estimation accuracy drops as $p$ increases. When $p = 50$ and $n = 30$, the TPR, FPR and MCC before and after the changepoint are (0.6, 0.01, 0.59) and (0.57, 0.03, 0.51) respectively. The computation complexity of the proposed MCMC algorithm increases as $p\times K$ increases. When $K = 5$, it took 0.8 minutes per 10000 iterations to run a one-changepoint model when $p\times K = 75$, 2.7 minutes when $p\times K = 100$, 6 minutes when $p\times K = 150$, 13 minutes when $p\times K = 200$ and 27 minutes when $p \times K= 250$. In general, the computational cost of our proposed model is dominated by the number of changepoints and the graphs size in the coefficient space.

\section{Application study}\label{sec::fgm_application}

\subsection{Sea surface temperature data}

Sea surface temperatures (SSTs) are a well-studied variable due to their importance to dynamical ocean circulations and to their effect on global weather, temperature, and precipitation.
SSTs vary at multiple and spatial time scales, including not only the diurnal and annual cycles but also a number of quasi-periodic or regime-like modes of variability.
For example, the El Niño Southern Oscillation (ENSO) is the leading mode of interannual climate variability and strongly modulates weather around the world \citep{Zebiak:1987cl,Ropelewski:1987do,Castillo:2014ci}.
Other modes of variability including the Pacific Decadal Oscillation \citep{Mantua:1997kj,newman:2016} and the the Madden-Julien Oscillation \citep{anderson_mjo:2020} affect climate at subseasonal to decadal time scales. Thus, SSTs demand flexible and dynamic models. 

\begin{figure}
    \centering
    \includegraphics[width=4in,height= 2in]{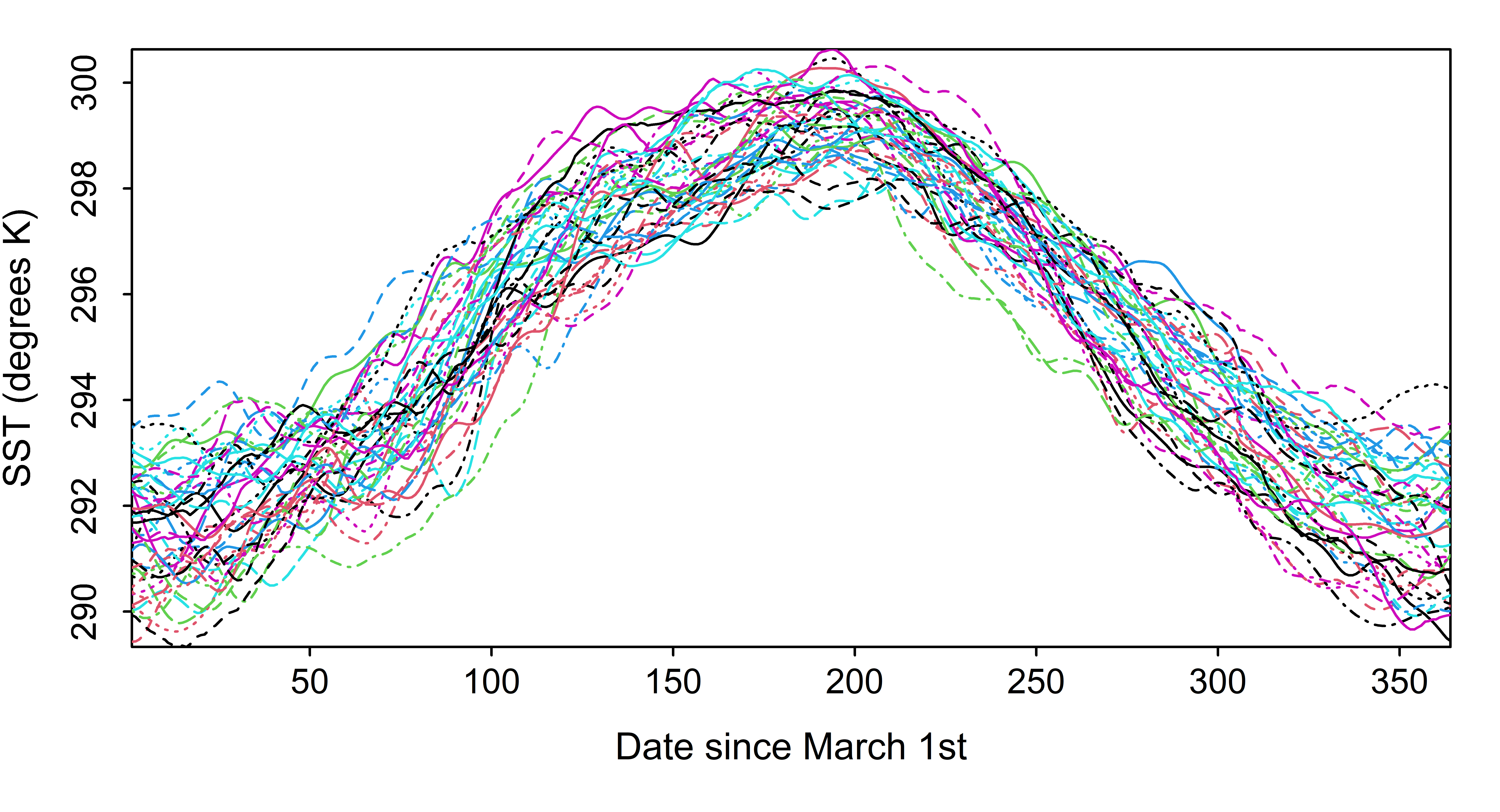}
    \caption{Observed SSTs $y_{i,j}(t)$ across day-of-year $t$ and years $i=1,\ldots,n$ (1979 - 2022) at location $j$ (Longitude = -160, Latitude= 30).}
    \label{sst_data}
\end{figure}

Here, we study the spatial relationships of SSTs in the tropical Pacific. Our goal is to analyze how temperatures at different locations influence each other and whether these influences change over the course of the year.
We use data from the ERA5 reanalysis, which uses a sophisticated physical model to assimilate satellite and in situ measurements of the global climate, providing a state-of-the-art reconstruction of past conditions \citep{hersbach_era5:2020}, from 1979 to 2022, defining each year to begin on March 1st. To model the complex and seasonal dynamics in the SSTs, we treat these measurements as functional data $y_{i,j}(t)$, where $t = 1,\ldots,T=365$ is the day-of-year, $i=1,\ldots,n=43$ is the year, and $j=1,\ldots,p$ are spatial locations. Figure \ref{sst_data} shows the functional data representation for a single location. Notably, the data are seasonal and relatively smooth, which suggests that the functional data models \eqref{truncated-model} and \eqref{truncated-model-cp} are appropriate. We choose $p=16$ grid cells (see Figure~\ref{sst_graph_static_sparse}) in the tropical Pacific and model these observations as multivariate functional data. We emphasize that the functional data model \eqref{truncated-model} is fully capable of modeling nonlinearities and seasonality, while the multivariate model \eqref{normal-prior} can adapt to the spatial dependencies among the curves. Despite the abundance of alternative space-time models \citep{cressie2015statistics}, these approaches are generally not designed for graph estimation. 

\subsection{Parameter settings}
We applied the proposed dynamic Bayesian functional graphical model to study the dependencies between annual SST functions at different locations. 
As demonstrated in Figure \ref{sst_data}, the data within one year are not strongly periodic. 
We therefore use a B-spline basis of order 3 (instead of a Fourier basis).
The model parameters are set to $\nu_0 = 0.02$, $h = 50$, $\alpha = 10$, $\beta = 40$ and $K = 10$. Model convergence was verified via the Geweke test on $\{\bm c^{i(s)}, \sigma_\epsilon^{(s)}, \tau\}$ with a significance level of 1\%. 
The choices of $\alpha$ and $\beta$ are different from those in the simulation study ($\alpha = 2$, $\beta = 7$), since those choices generated very sparse graphs for this application (see Appendix~\ref{sec::sensitivity-study-application}). Therefore, we increased $\alpha$ and $\beta$ to strengthen the prior and elevate the posterior block-wise edge inclusion probabilities. 
In the simulation study, the values $\alpha = 10$ and $\beta = 40$ returned results almost identical to those reported in Section~\ref{sec::simulation-results} (see Appendix~\ref{sec::sensitivity-study-application}). The number of basis functions $K$ was chosen based on DIC, as defined in \cite{gelman1995bayesian}. For $K$ = (5, 8, 10, 12), we obtained DIC = (-59095, -135343, -177387, -76873) and therefore set $K$ to 10.

\subsection{Results}
\begin{figure}[H]
    \centering
    \includegraphics[width=.75\textwidth]{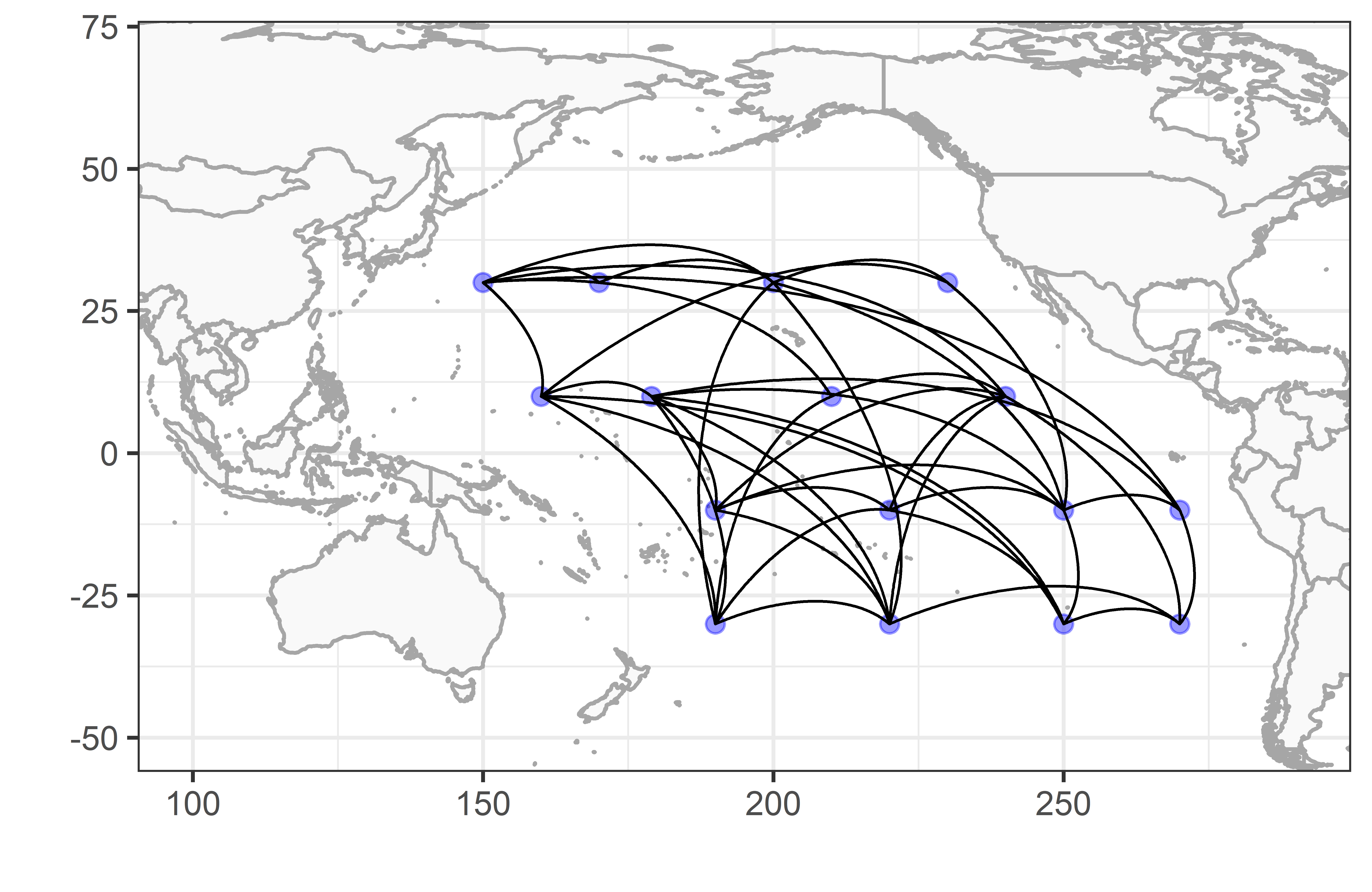}
            \caption{Graph estimation from the static Bayesian functional graphical model on the sea surface temperature data.}
    \label{sst_graph_static_sparse}
\end{figure}

First, we show results from the static Bayesian functional graph described in Section~\ref{bfgm}. The estimated graph is presented in Figure~\ref{sst_graph_static_sparse}. There are 82 edges, which is 35\% of the total number of possible edges.
The locations in the middle of the Pacific ocean are unconnected (or conditionally independent) along the longitude direction.
The majority of vertical communications are located near the coastal lines. Almost all grid locations are connected to points to the East and West; very few nodes are connect to locations to their North or South.
These findings are consistent with zonally dominated currents in the study region.
In addition, edges from non-adjacent grid cells, i.e. teleconnections, are apparent between the nodes in the domain's Northwestern and Southeastern corners.
We notice that existing studies yield graphs that are densely or even fully connected in this region \citep{tsonis2004architecture, tsonis2006networks}.
One possible explanation is that our model may better identify the underlying dynamic mechanism, and in general graphical models are well-suited to omit spurious correlations. 
A detailed examination of the edges in climate sciences is beyond the scope of this paper. However, we emphasize that a sparse graph is a more appealing and parsimonious explanation of the CN, especially compared to traditional methods, which typically build a densely-connected graph on a fine grid then describe the graph through topological summaries \citep{donges2009complex, fan2021statistical}.  
   
\begin{figure}[H]
    \centering
    \includegraphics[width=.49\textwidth]{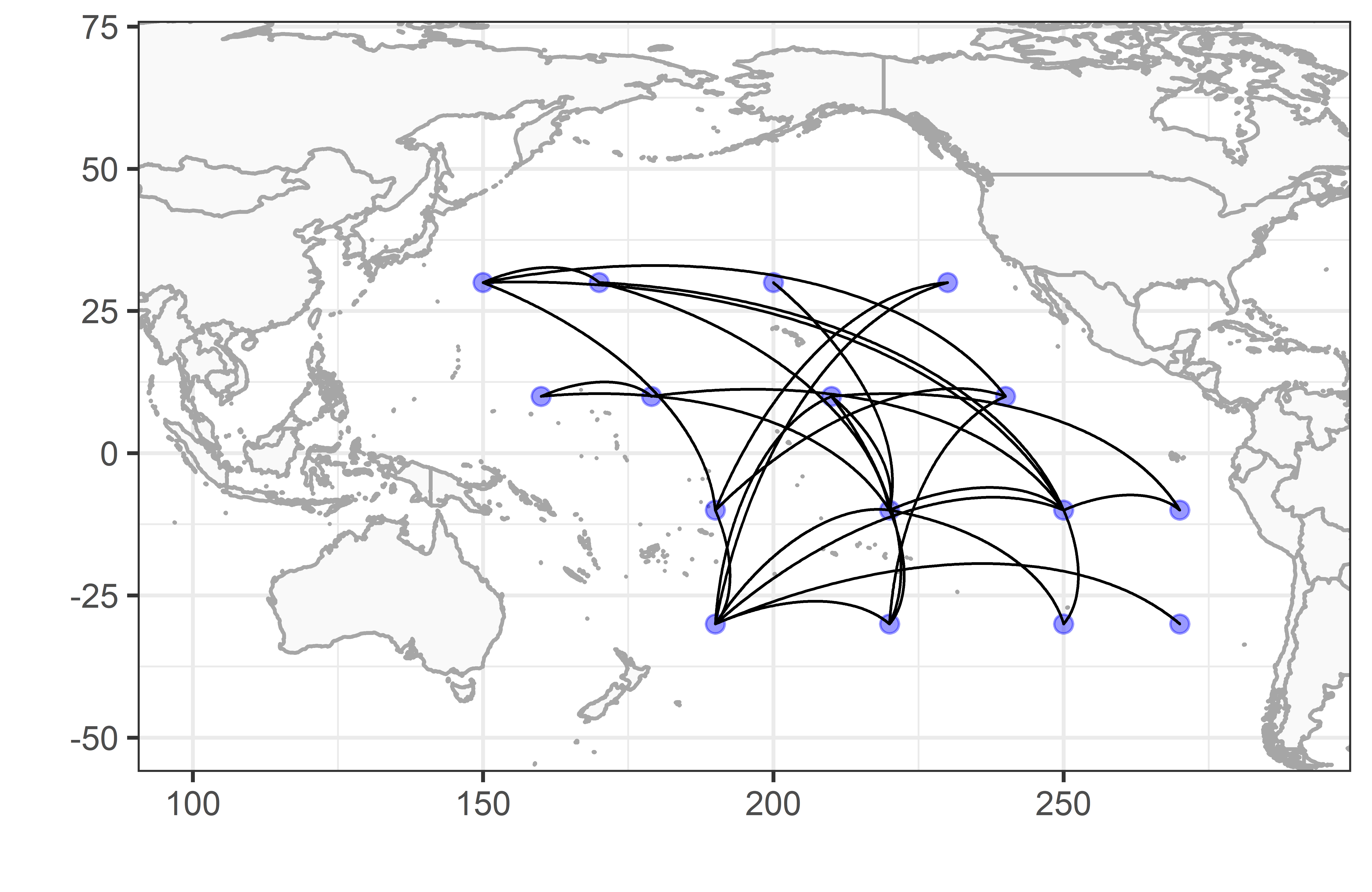}
    \includegraphics[width=.49\textwidth]{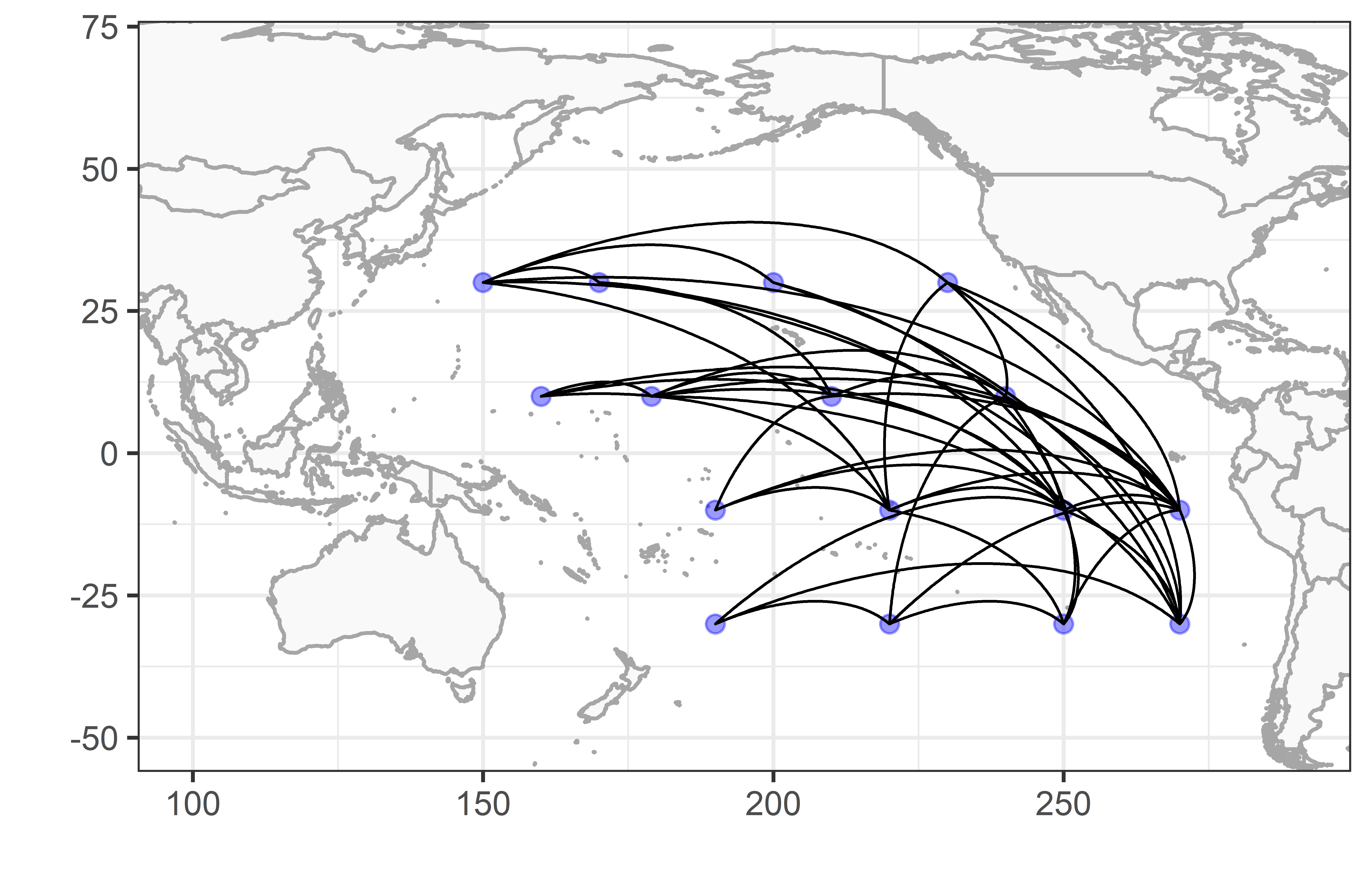}
    \caption{Graph estimation before the changepoint (left) and after the changepoint (right) on the sea surface temperature data.}
    \label{sst_graph_dynamic}
\end{figure}

Next, we fit the proposed dynamic Bayesian functional graphical model with one changepoint. The posterior distribution of $\tau$ concentrated around August 29th ($t = 181$), even as we vary the changepoint initializations in the MCMC sampling algorithm. 
The estimated graphs pre-changepoint (March 1 - August 29) and post-changepoint (August 30 - February 28) are plotted in Figure~\ref{sst_graph_dynamic}.
From March to August, most of the zonal (East-West) connections appear on the Western side of the plot; from September to February, the Eastern side is more meridionally (North-South) connected.
More meridional edges are found during fall and winter.
Similar to results for the static model, there are more zonal connections in both graphs.
In addition, more teleconnections from the North West to the South East are observed during fall and winter.
In conclusion, the static model is a summary of all edges in both of these graphs;  the dynamic model reveals the structural change of the graph at the changepoint, with a noticeably sparser graph pre-changepoint. 

Last, we expand the dynamic graph to include three changepoints, which partition the year into four subdomains (i.e., seasons), respectively. The edges are mostly consistent with those detected in the one-changepoint model, and the model yields an DIC of -68887, which is greater than the one-changepoint scenario. Therefore, the functional graphs estimated in Figures~\ref{sst_graph_static_sparse}~and~\ref{sst_graph_dynamic} are most appropriate. Details of the results are included in Appendix \ref{sec::sensitivity-study-application}.

\section{Conclusions} \label{sec:fgm_conclusions}

We introduced a Bayesian graphical modeling framework to learn the conditional dependencies in multivariate functional data. By pairing a basis expansion with a block-structured sparsity prior, the proposed approach enables flexible modeling of complex functional (or dynamic) patterns along with powerful graph estimation. The model further incorporates a changepoint to identify both time-invariant and time-varying connectivity patterns, and unlike existing approaches, defines a \emph{dynamic} functional graph that is consistent with the traditional definition of a functional graph. 
Compared to static functional graphs, which incorporate information over the entire functional domain and thus are typically more dense, our dynamic functional graphs often provide simpler and more parsimonious explanations. Simulation studies confirmed the effectiveness of our approach, which provided more accurate graph estimation and more powerful edge detection compared to state-of-the-art frequentist and Bayesian competitors. We applied both static and dynamic versions of our model to sea surface temperature data and discovered edges with practical meanings. 

Although the sea surface temperature data could be modeled accurately using a relatively small number of basis functions ($K=10$), other applications may require larger values of $K$. However, increasing $K$ significantly increases the computational burden, since the relevant precision matrices are $(pK)\times (pK)$. In general, the computational cost of our proposed model is dominated by the number of graphs and the graph size $(pK)\times (pK)$. Alternatively, neighborhood selection approaches may provide more scalability for these  settings \citep{zhao2021high}, and similarly could be extended to incorporate (one or more) changepoints for dynamic graph estimation. The model performance benefits from estimating each edge in $\bm G$ separately using marginal edge inclusion probabilities, which is standard in Bayesian graphical models.

There are several promising avenues for future work. First, there are alternative approaches to construct a posterior estimate of a graph, such as fitting a graphical lasso to the posterior mean of the precision matrix. Second, regime-switching behavior of the graph is often interesting, and our static functional graphical model could be used as the emission distribution of a hidden Markov model \citep{warnick2018bayesian,liu2022dynamic}. Third, instead of DIC-based selection of the number of changepoints, we could incorporate a prior on the number of changepoints \citep{franzolini2024change}. Fourth, extensions are available to non-Gaussian data via copula graphical models. Finally, while the proposed model assumes that the coefficients are conditionally independent pre- and post-changepoint, dependencies among coefficients could be incorporated. 

\section*{Code availability}
Code implementing the model described in this paper can be downloaded from GitHub at 
https://github.com/chunshanl/DBFGM (to be made public at acceptance).

\section*{Statements and declarations}
The authors have no relevant financial or non-financial interests to disclose. The authors have no competing interests to declare that are relevant to the content of this article. All authors certify that they have no affiliations with or involvement in any organization or entity with any financial interest or non-financial interest in the subject matter or materials discussed in this manuscript. The authors have no financial or proprietary interests in any material discussed in this article.

\bibliographystyle{apalike}
\bibliography{refs}

\newpage

\begin{appendices}
\section*{Appendix}

\section{Details of the MCMC algorithm} \label{sec::mcmc_dbfgm}

Here, we detail the steps of our proposed MCMC algorithm. For generality, we suppose that there are $S-1$ changepoints $\{\tau^{(1)}, \ldots, \tau^{(S-1)}\}$ that can take values in \{1, 2, \ldots, T\}. Setting $\tau^{(0)} = 1$ and $\tau^{(S)} = T$, these changepoints partition the functional domain $\mathcal{T}$ into subdomains $T^{(1)},\ldots, T^{(S)}$ with $T^{(s)} = (\tau^{(s-1)}, \tau^{(s)}]$. Then for replicates $i=1,\ldots,n$, functions $j=1,\ldots,p$, and states $s=1,\ldots,S$, the full model is
\begin{align}
\begin{split}
    Y^{i}_{j}(t) &= 
    \sum_{k=1}^{K} c^{i(s)}_{j,k} f_{k}(t) + \sigma_\epsilon^{(s)} \epsilon^{i}_{j}(t), \quad \epsilon^{i}_{j}(t) \stackrel{iid}{\sim}N(0,1),\quad t\in T^{(s)} \\
    [\bm c^{i(s)}\mid \bm \Omega^{(s)}] &\stackrel{indep}{\sim} 
    MVN(\bm 0, [\bm\Omega^{(s)}]^{-1}), \\
    p(\bm \Omega^{(s)}\mid  \bm G^{(s)},...) 
    & =
    C(\bm G^{(s)}, v_0, v_1, \lambda)^{-1} \cdot \\
    &  \prod_{j_1< j_2, k_1, k_2} 
    N\{ \omega^{(s)}_{j_1 k_1, j_2 k_2} \mid  0, 
      v_0^2 + 
    g^{(s)}_{j_1 k_1, j_2 k_2} \cdot (v_1^2 - v_0^2)
    \} \cdot \\
    & \prod_{j,1 \leq k_1 < k_2 \leq K} 
    N\{ \omega^{(s)}_{j k_1, j k_2} \mid  0, 
      v_0^2 + 
    g^{(s)}_{j k_1, j k_2} \cdot (v_1^2 - v_0^2)
    \} \cdot \\
    & \prod_{j,k} 
    \textrm{Exp}( \omega^{(s)}_{jk, jk} \mid  \lambda/2) \cdot
    \bm 1_{\bm \Omega^{(s)} \in M^{+}}, \\
    p(\bm G^{(s)} \mid \bm \pi^{(s)}, ...) 
    & =
    C(\bm \pi^{(s)}, v_0, v_1, \lambda)^{-1} \cdot
     C(\bm G^{(s)}, v_0, v_1, \lambda) \\
    & \prod_{j_1<j_2} 
     \prod_{k_1, k_2} 
     \pi_{j_1 j_2}^{(s) g_{j_1 k_1, j_2 k_2}} (1-\pi_{j_1 j_2}^{(s)})^{(1-g_{j_1 k_1, j_2 k_2})}\\
    & \prod_{j} 
     \prod_{k_1<k_2} 
     \pi_{0}^{(s) g_{j_1 k_1, j_2 k_2}} (1-\pi_{0}^{(s)})^{(1-g_{j k_1, j k_2})}, \\
         p(\bm \pi^{(s)}) &=  C(v_0, v_1, \lambda)^{-1} C(\bm \pi^{(s)}, v_0, v_1, \lambda)  \textrm{Beta}(\pi^{(s)}_0\mid  \alpha_0, \beta_0) 
    \prod_{j_1 < j_2} \textrm{Beta}(\pi^{(s)}_{j_1, j_2}\mid \alpha, \beta), \\ 
    \sigma^{(s)2}_{\epsilon} & \stackrel{iid}{\sim} \textrm{Inverse-Gamma}(\alpha_\sigma, \beta_\sigma),\\
    p(\tau^{(s)}\mid  \tau^{(r)}, \forall r \neq s) & = \textrm{Uniform}(\{\tau^{(s-1)}, \tau^{(s-1)} + 1, \ldots, \tau^{(s+1)} \})
\end{split}
\label{full-prior}
\end{align}
where $C(\cdots)$ denotes a normalizing constant that depends only on the terms in the parentheses. 

The posterior sampling algorithm is as follows: 
\begin{itemize}

\item \textbf{Update ${\bm \Omega^{(s)}}$:}  
Following \cite{peterson2020bayesian}, for $s \in \{1,..,S\}$, sample ${\bm \Omega}^{(s)}$ one column and row at a time. Here we demonstrate how to sample the last row and last column for $\bm \Omega^{(s)}$. First, partition $\bm\Omega^{(s)}$ into 
\begin{equation*}
    \bm\Omega^{(s)} = \begin{pmatrix}
\bm \Omega^{(s)}_{11} & \bm \omega^{(s)}_{12}  \\
\bm \omega^{(s)}_{21} & \omega^{(s)}_{qq}  
\end{pmatrix} 
\end{equation*}
where $q = pK$, $\omega^{(s)}_{qq}$ is the $(q,q)$th element of the matrix, $\bm \omega^{(s)}_{12}$ is the last column except $\omega^{(s)}_{qq}$ and $\bm \omega^{(s)}_{21}$ is the last row except $\omega^{(s)}_{qq}$. Similarly, partition $\bm\Sigma^{(s)} = \sum_{i=1}^{n} \bm c^{i(s)} \bm c^{i(s)'} $ into
\begin{equation*}
\bm \Sigma^{(s)} = \begin{pmatrix}
\bm \Sigma^{(s)}{11} & \bm \sigma^{(s)}_{12}  \\
\bm \sigma^{(s)}_{21} & \sigma^{(s)}_{qq}  
\end{pmatrix} .
\end{equation*}

Next, perform a one-on-one transformation of the last column of $\bm \Omega^{(s)}$ as $(\bm a , b) = (\bm \omega^{(s)}_{12}, 
\omega^{(s)}_{qq} - \bm \omega^{(s)}_{21} (\bm\Omega^{(s)}_{11})^{-1} \bm \omega^{(s)}_{12} )$. It can be proved that the conditional posterior of $(\bm a, b)$ is 
\begin{equation}
\begin{aligned}
        p(\bm a, b \mid \ldots)
    =
    \textrm{Normal}(\bm Q^{-1} \bm l, \bm Q^{-1}) \textrm{Gamma}(\dfrac{|T^{(s)}|}{2}+1, \dfrac{\sigma^{(s)}_{qq}+\lambda}{2})
\end{aligned}
\end{equation}
where 
$\bm l = - \bm \sigma^{(s)}_{12}$, 
$\bm Q = diag(\bm v^{(s)}_{12})^{-1}  + (\sigma^{(s)}_{qq}+\lambda) (\bm\Omega^{(s)}_{11})^{-1}$, and the second parameter in the gamma distribution is rate. Here $\bm v_{12}^{(s)}$ are the variances of $\bm \omega^{(s)}_{12}$, which takes values in $v_0^2$ and $v_1^2$ given the edge inclusion indicators $\bm G^{(s)}$. If $T^{(s)}$ is empty, we set $\bm \Sigma^{(s)} = \bm 0$ and sample the precision matrix from its prior.

The positive definiteness of $\bm \Omega^{(s)}$ is automatically guaranteed by the Sylvester’s criterion. Assume the sampled $\bm \Omega^{(s)}$ at the $m$th iteration, denoted as $\bm \Omega^{(s)(m)}$, is positive definite. After updating its last column and row at the $(m+1)$th iteration, the leading $q-1$ principal minors of $\bm \Omega^{(s)(m+1)}$ are still positive. Note that the principal minor of $q$th order in $\bm \Omega^{(s)(m+1)}$ is $b \cdot |\bm \Omega_{11}^{(s)(m)}|$. Since $b$ is positive with a Gamma distribution, the $q$th leading principal minor is positive and $\bm \Omega^{(s)(m+1)}$ is positive definite.
 
\item \textbf{Update $\bm G^{(s)}$:}  Sample $\bm G^{(s)}$, $s=1,\ldots,S$ from the conditional Bernoulli distributions
\begin{equation*}
    p(g^{(s)}_{i_1 i_2}=1\mid \ldots)=
\dfrac{N(\omega^{(s)}_{i_1 i_2}\mid 0,v_1^2)\pi_{i_1 i_2}^{(s)}}
{N(\omega^{(s)}_{i_1 i_2}\mid 0,v_1^2)\pi_{i_1 i_2}^{(s)}+N(\omega^{(s)}_{i_1 i_2}\mid 0,v_0^2)(1-\pi_{i_1 i_2}^{(s)})}.
\end{equation*}
Here $i_1,i_2$ indicates all the off-diagonal elements in the $pK \times pK$ matrix, and the edge inclusion probabilities $\pi_{i_1 i_2}^{(s)}$ are determined by the block-wise edge inclusion probabilities $\bm \pi^{(s)} = (\pi^{(s)}_0, \{\pi^{(s)}_{j_1 j_2}\}_{ 1 \leq j_1<j_2 \leq p} )$.

\item{Update $\bm \pi^{(s)} =  (\pi^{(s)}_0, \{\pi^{(s)}_{j_1 j_2}\}_{ 1 \leq j_1<j_2 \leq p} )$, $s = 1,\ldots, S$:} Given graph $\bm G^{(s)}$, the posterior of each element in $\bm \pi^{(s)}$ is a conjugate beta distribution with
\begin{equation*}
    p(\pi^{(s)}_0 \mid ...) = \textrm{Beta}
    (\alpha_0  + \sum_j \sum_{k_1<k_2} g^{(s)}_{jk_1, jk_2}, \beta_0 + K(K+1)/2 \cdot p - \sum_j \sum_{k_1<k_2} g^{(s)}_{jk_1, jk_2})  
\end{equation*}

\begin{equation*}
    p(\pi_{j_1j_2}^{(s)} \mid ...) = 
    \textrm{Beta}
    (\alpha  + \sum_{k_1, k_2} g^{(s)}_{j_1k_1, j_2k_2}, \beta + K^2 -  \sum_{k_1, k_2} g^{(s)}_{j_1k_1, j_2k_2})
\end{equation*}

\item{Update $\bm c^{i(s)}$, $s = 1,\ldots, S$, $i = 1, \ldots, n$:} For each $i$ and $s$, sample the basis coefficients from 
$MVN(\bm c^{i(s)} \mid \bm \mu_i^{s}, \bm \Sigma_i^{s}))$
where $\bm\mu_i^{s} = \bm Q^{-1} \bm{l}$, $ \bm \Sigma_i^{s}
 = \bm Q^{-1} $, with
\begin{align*}
    & \bm Q = \sigma_{\epsilon}^{(s)-2}  \sum_{t \in T^{(s)}}
    \bm F(t)'\bm F(t) + \bm \Omega^{(s)}\\
    & \bm l = \sigma_{\epsilon}^{(s)-2} 
\sum_{t \in T^{(s)}} \bm F(t)' \bm y^i(t) 
\end{align*}

where $\bm F(t) $ is defined as  
$$
 \bm F(t) = 
 \begin{pmatrix}
 \bm f_1'(t) & 0 & 0 & \hdots & 0 \\
0&   \bm f_2'(t)  & 0 & \hdots & 0 \\
\vdots & & \ddots & \hdots & \vdots \\
0 &   & &  0 &  \bm f_p'(t) 
 \end{pmatrix}_{p \times pk}
$$
and $\bm f_j'(t) = (f_{1}(t), \ldots, f_{K}(t))$. If $T^{(s)}$ is empty, set $\bm l = \bm 0$ and $\bm Q = \bm \Omega^{(s)}$.

\item{Update $\sigma_{\epsilon}^{(s)}$:} For each $s$, sample 
\[
\sigma_\epsilon^{(s)2} \sim \textrm{Inverse-Gamma}(\alpha_\sigma + \frac{1}{2}|T^{(s)}|np, \beta_\sigma + \frac{1}{2} \sum_{t \in T^{(s)}} \sum_{i, j} \{Y^{i}_{j}(t) -
    \bm f'_j(t) \bm c^{i(s)}_{j}\}^2)
\]
    
\item{Update $\tau^{(s)}$}, $s = 1, \ldots S-1$:  The posterior of the changepoint $\tau$ is a discrete distribution on $\{\tau^{(s-1)}, \tau^{(s-1)} + 1, \ldots, \tau^{(s+1)}\}$ with weights 
\begin{align*}
    p(\tau^{(s)} = \tau \mid \tau^{(r)}, \forall r \neq s, \ldots) 
    \propto \prod_i\prod_j 
    &\{ \prod_{\tau^{(s-1)} \leq t < \tau} N[Y^{i}_{j}(t) -
    \bm f'_j(t) \bm c^{i(s-1)}_{j}\mid  0, (\sigma_\epsilon^{(s-1)})^{2}] \\
    & \prod_{\tau \leq t < \tau^{(s+1)}}N[Y^{i}_{j}(t) -
    \bm f'_j(t) \bm c^{i(s)}_{j}\mid  0, (\sigma_\epsilon^{(s)})^{2}]   \}.
\end{align*}

To make the MCMC more computationally efficient, one could constrain a changepoint inside a compact interval $(t_1, t_2)$ based on domain knowledge from the application. 

\end{itemize}

\section{Supplementary material for simulation study}
\label{sec::prior-evaluation}

\begin{table}[H]
\footnotesize{
\centering
\begin{tabular}{|c| c c c |c c c| c c c |c c c |}  
\hline
 & $h=10$ & $h=50$ & $h = 100$ \\
 \hline
K = 5 & 0.2 & 0.13 & 0.07   \\
K = 8 & 0.19 & 0.1 & 0.04  \\
K = 10 & 0.18 & 0.07 & 0.03 \\
\hline
\end{tabular}               
\caption{Prior edge inclusion probability of graph $\bm{G}$ in the coefficient space using different values of $K$ and $h$. Parameters are set to $\alpha = 2$, $\beta=7$, $v_0 = 2$ and $p = 15$.}
\label{tab:prior-inclusion-table}
}
\end{table}

\begin{figure} [H]
\centering
\begin{tabular}{cccc}
\includegraphics[width=0.5\textwidth]{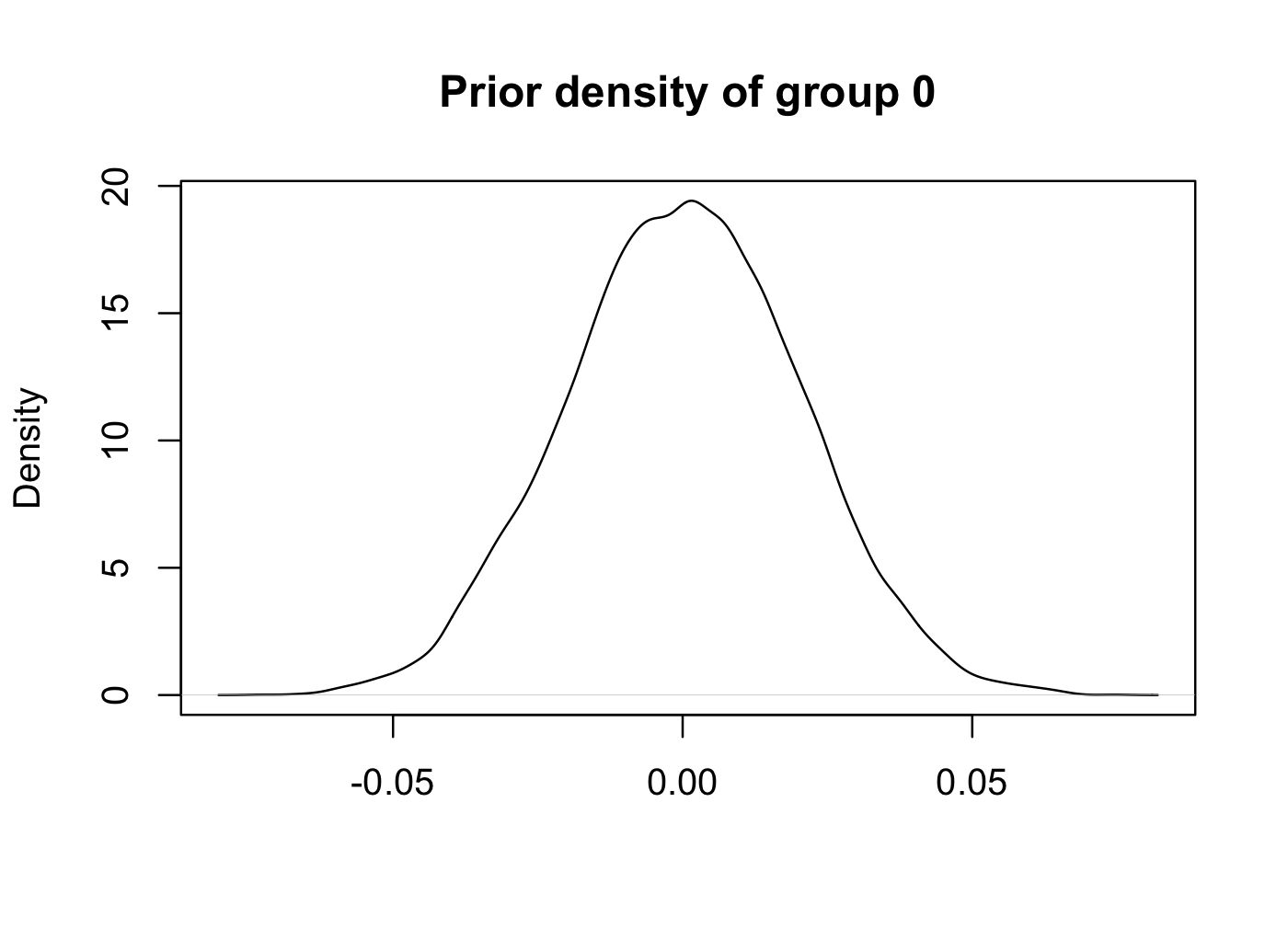}&
\includegraphics[width=0.5\textwidth]{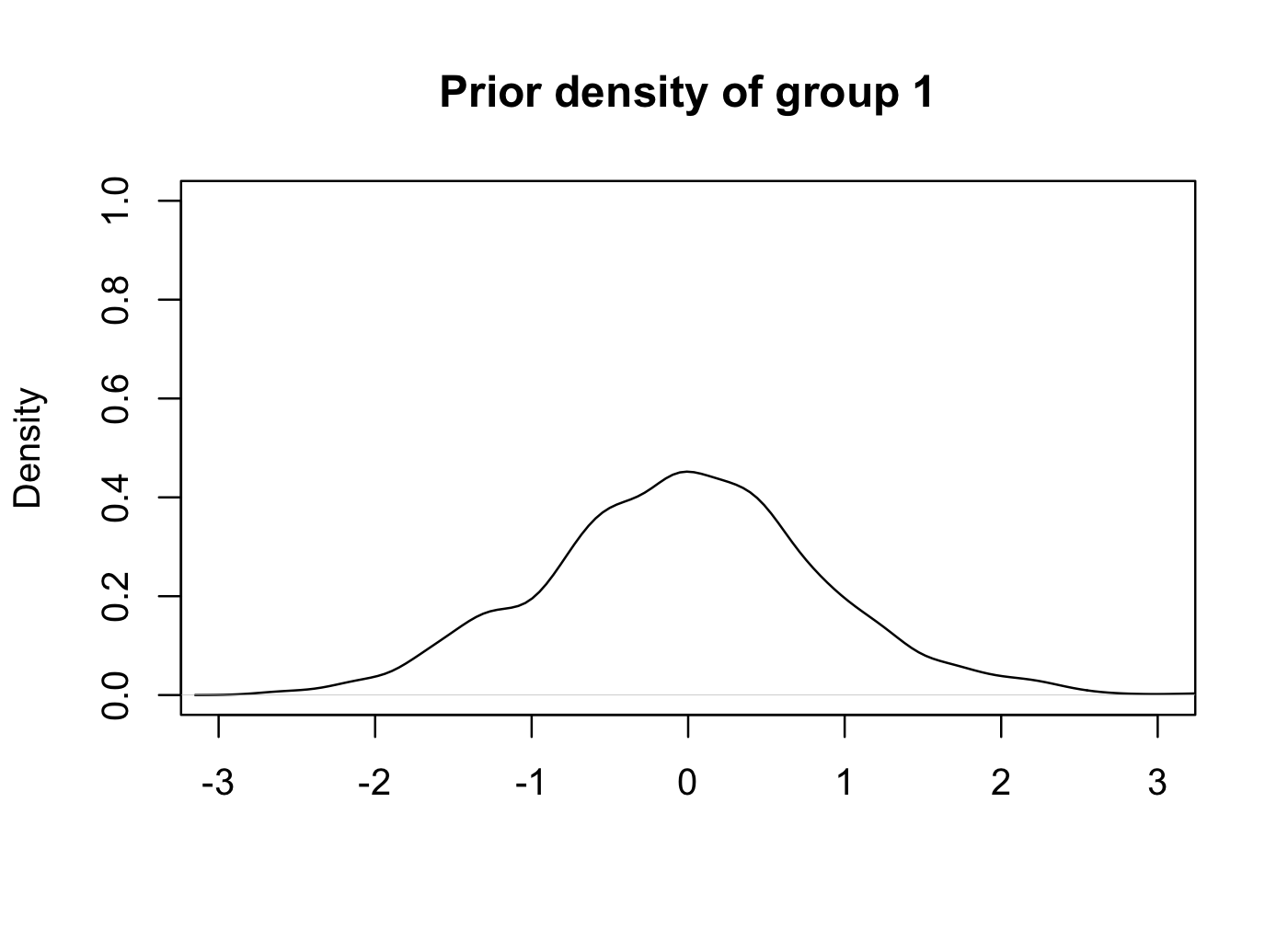} 
\end{tabular}
\caption{Prior distributions of an element in the precision matrix given that the edge is absent (left) or present (right). The scales of the x-axes are different.}
\label{precision-prior}
\end{figure}

\begin{figure} [H]
\centering
\begin{tabular}{cccc}
\includegraphics[width=0.5\textwidth]{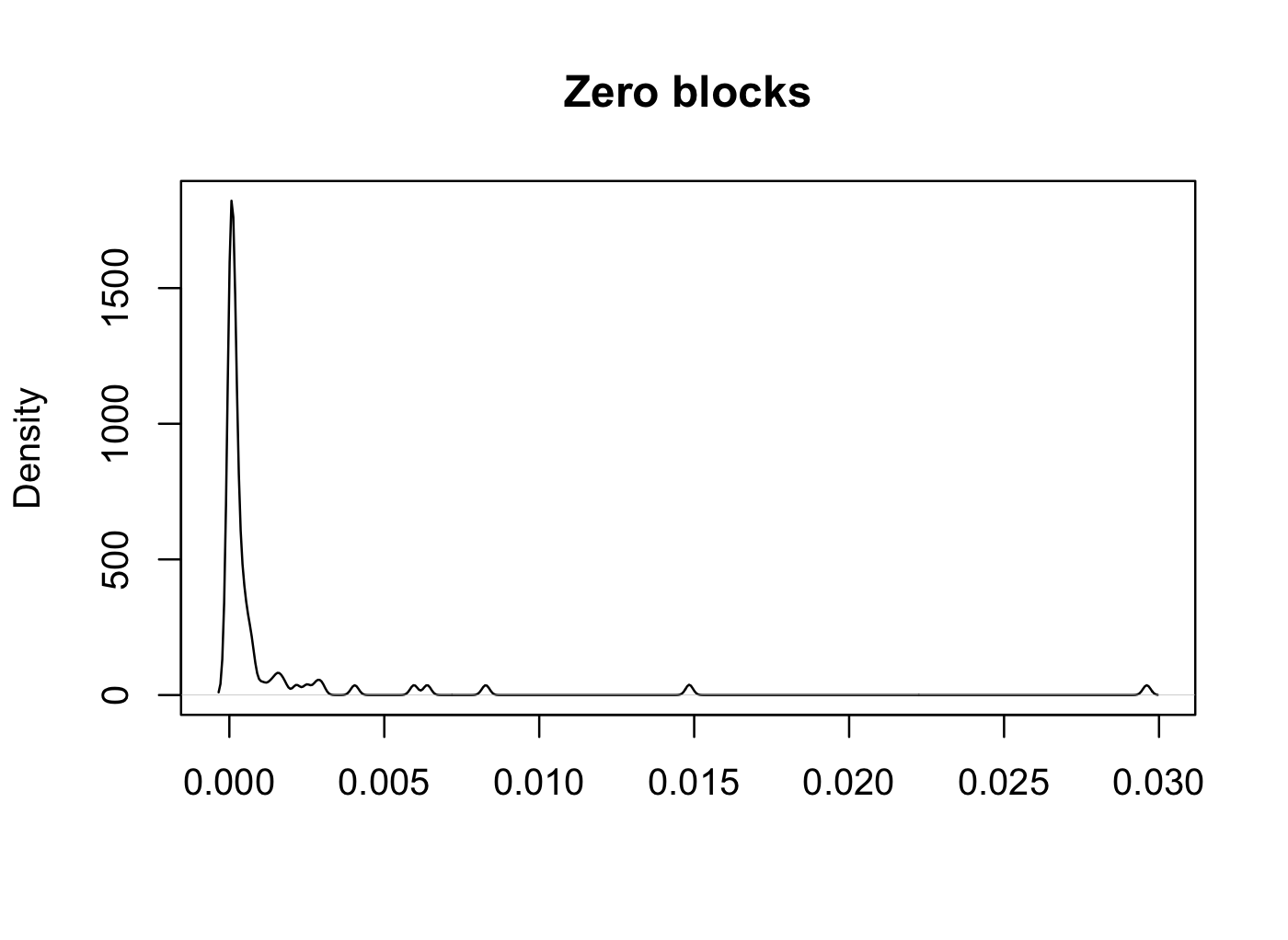}&
\includegraphics[width=0.5\textwidth]{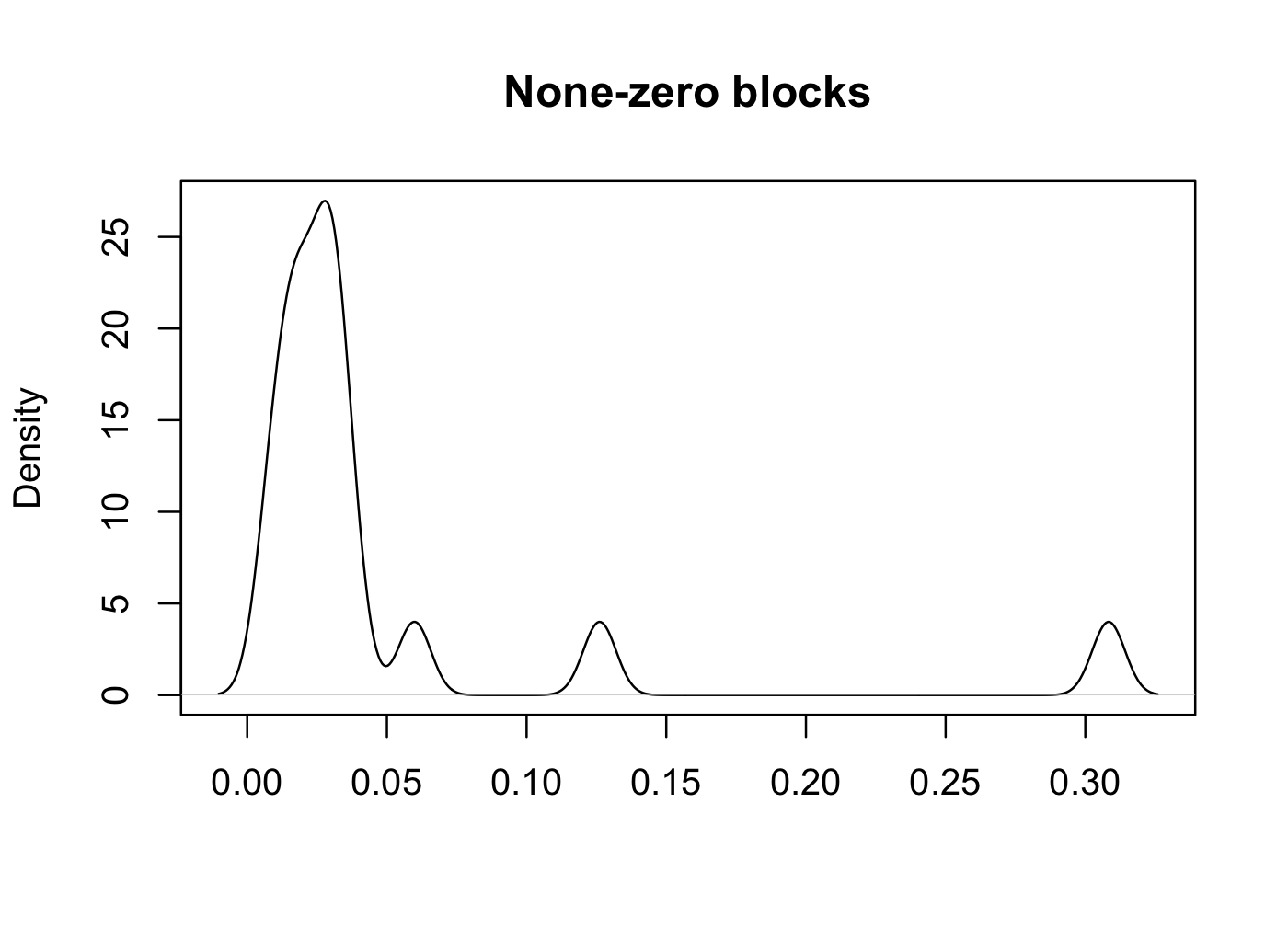} 
\end{tabular}
\caption{Distribution of posterior block-wise Frobenious norm of the precision matrix given edge estimation in the functional space. The left plot is when the corresponding edge in the functional space is estimated to be absent; the right plot is when the corresponding edge in the functional space is estimated to be present.}
\end{figure}

\begin{figure}[H]
\centering
    \includegraphics[width=0.4\textwidth]{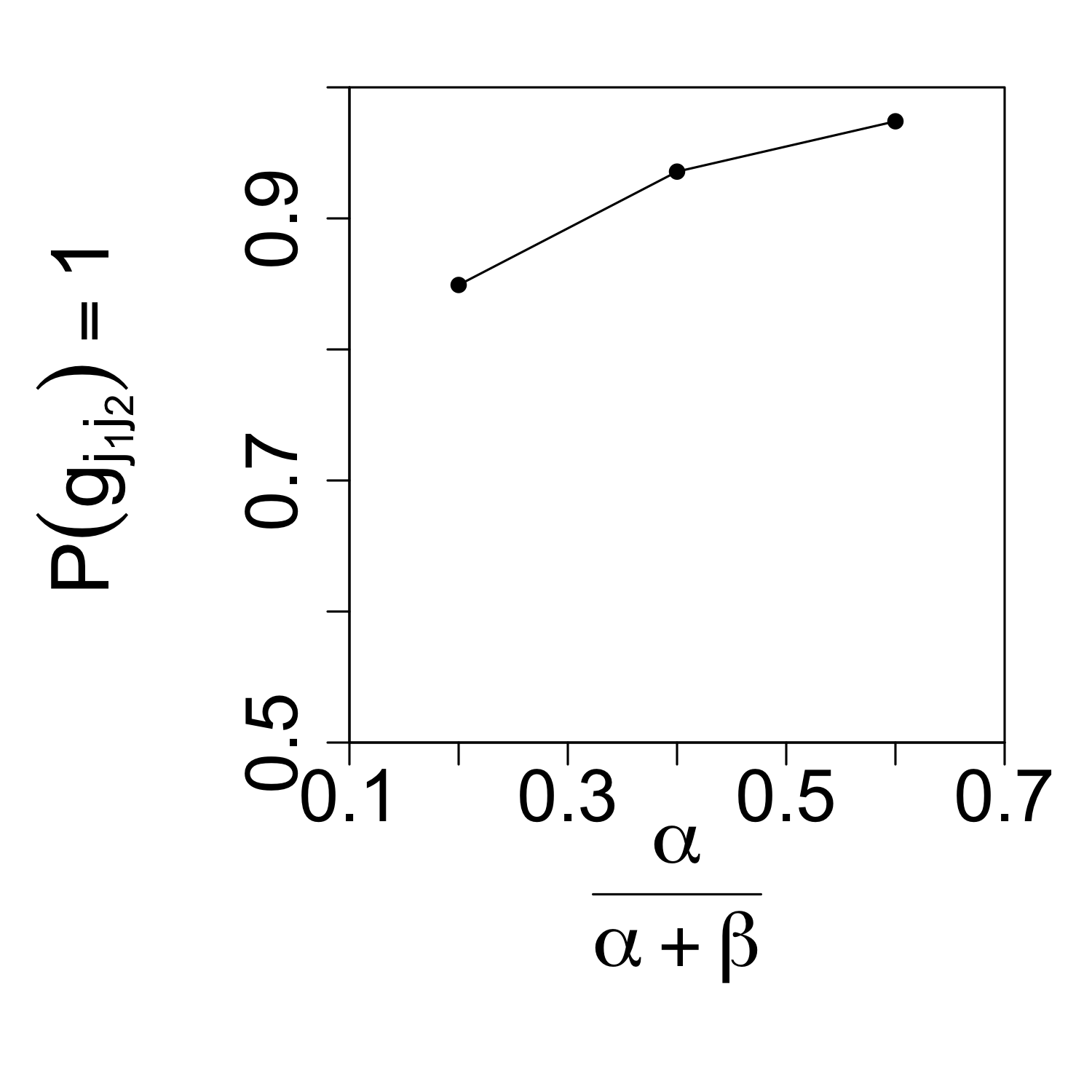}
    \includegraphics[width=0.4\textwidth]{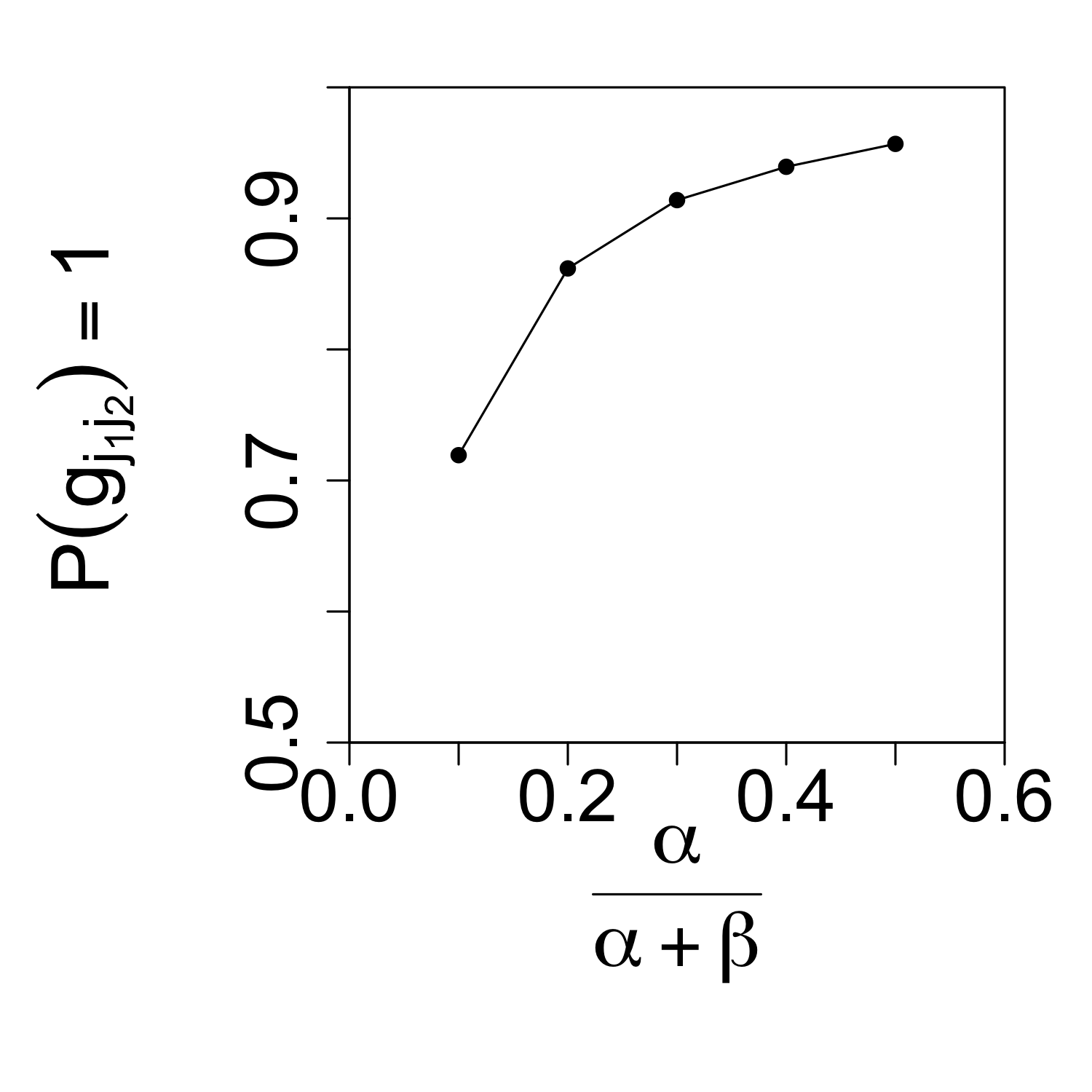} \\
    \includegraphics[width=0.4\textwidth]{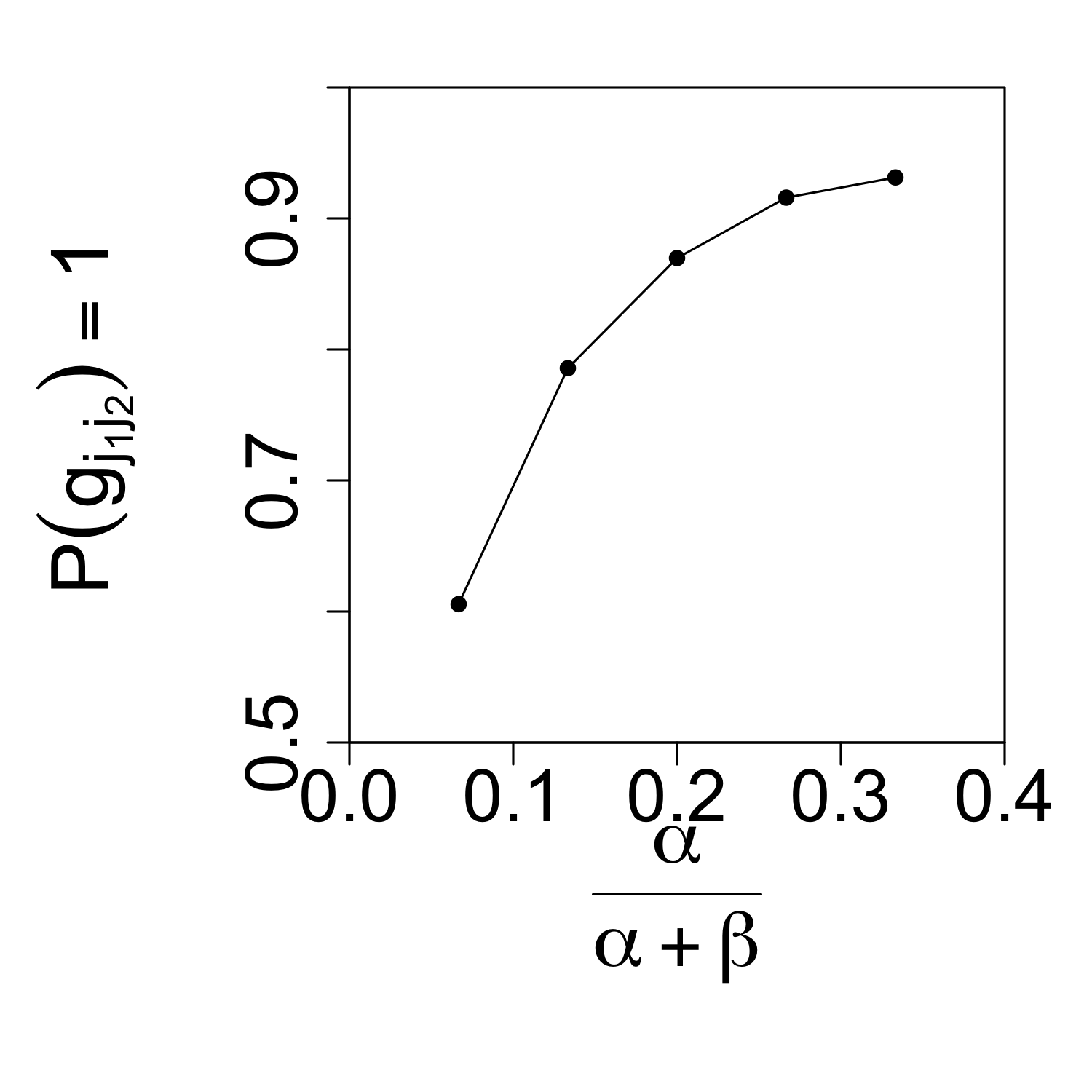} 
    \includegraphics[width=0.4\textwidth]{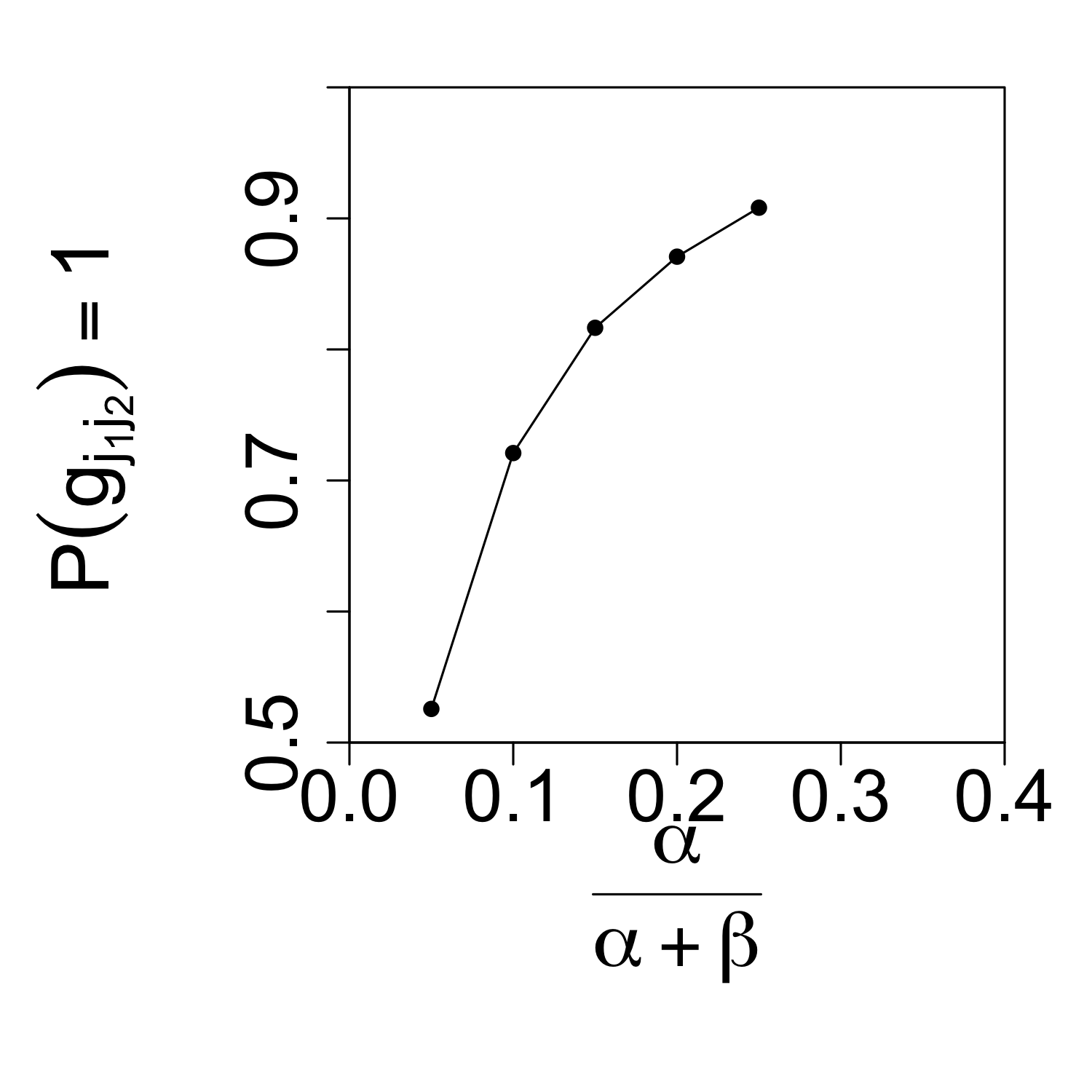}
        \caption{Prior edge inclusion probability for $\mathcal{G}$  (y-axis) versus mean of the Beta prior, $\alpha/(\alpha + \beta)$, (x-axis)  for $p=5$ (top left), $p=10$ (top right), $p=15$ (bottom left), and $p=20$ (bottom right). The edge inclusion probabilities are uniformly higher than for $\bm G$ (Figure~\ref{prior-inclusion-probability}).}
    \label{prior-inclusion-probability-alt}
\end{figure}


\section{Supplementary material for the application study}
\label{sec::sensitivity-study-application}

\begin{figure} [H]
\centering
\begin{tabular}{cccc}
\includegraphics[width=0.35\textwidth]{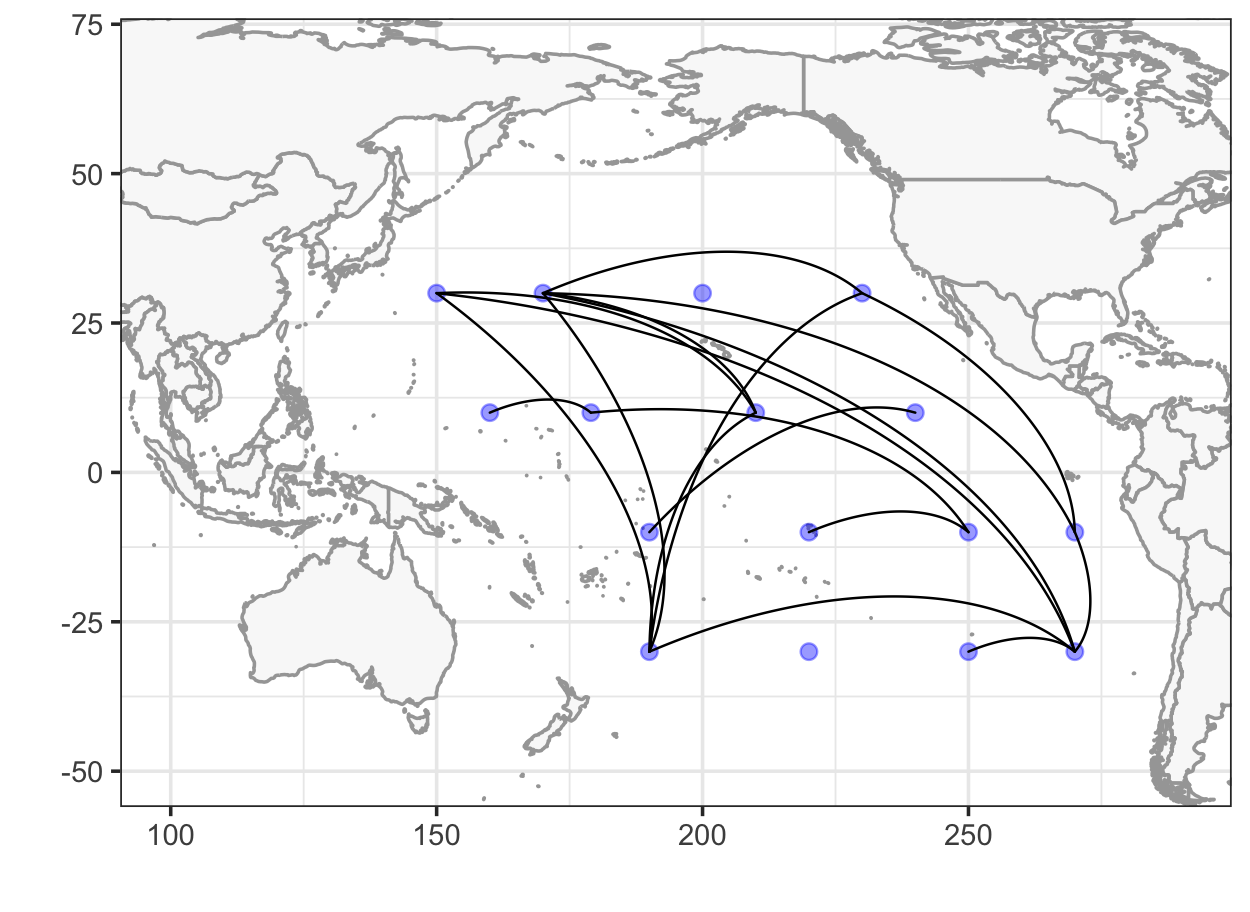}&
\includegraphics[width=0.35\textwidth]{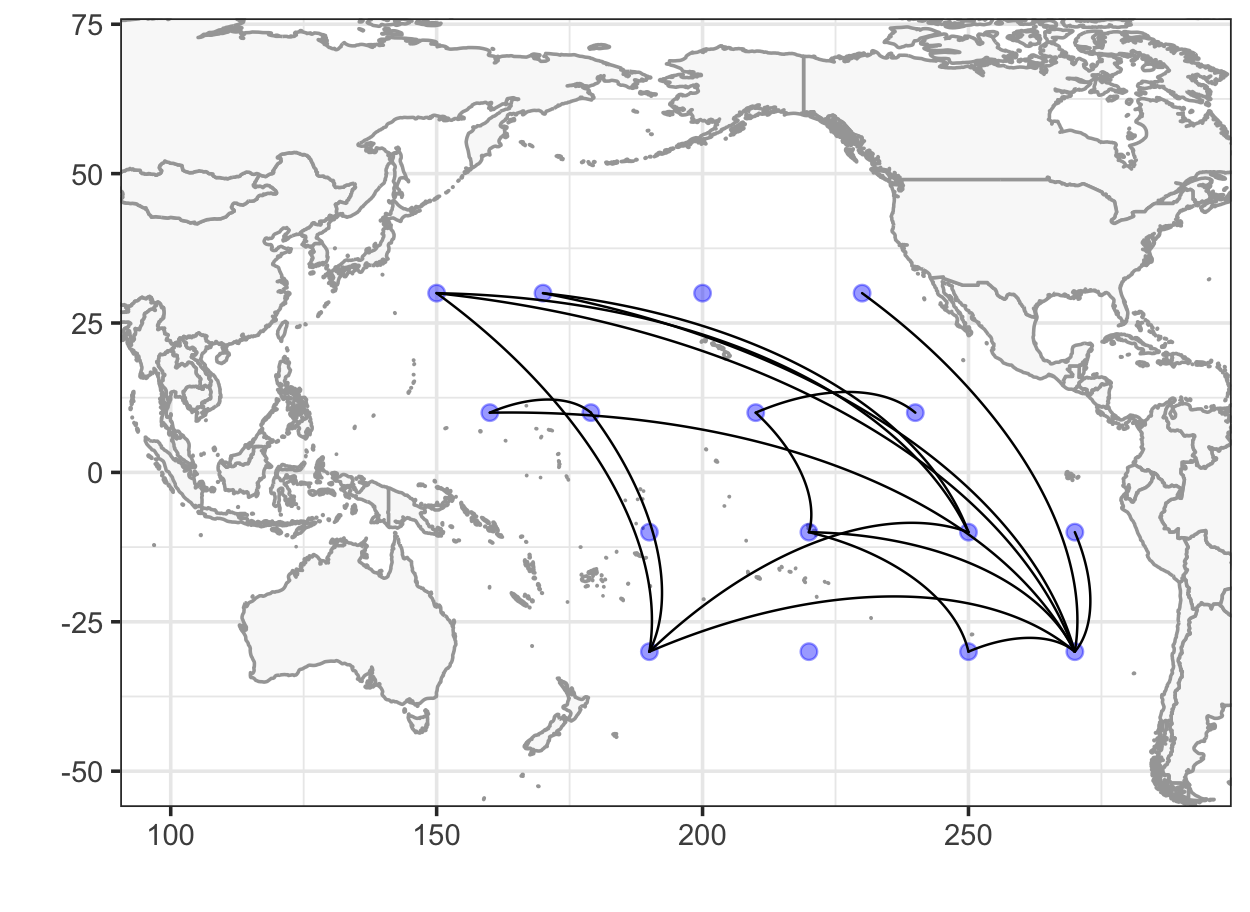} 
\end{tabular}
\caption{Posterior estimation of graphs before and after the changepoint when $\alpha = 2$ and $\beta = 7$.}
\label{sst-sparse}
\end{figure}

Varying the hyperparameters, we also fit the proposed one-changepoint model with $\alpha=2$ and $\beta = 7$. The resulting graphs are in Figure \ref{sst-sparse}. Since these graphs are exceedingly sparse, the main paper considers a stronger prior (i.e., a smaller prior variance) with larger values of $\alpha$ and $\beta$. For further clarification, consider the conditional posterior of $\pi_{j_1j_2}$ from the MCMC algorithm: 
\begin{equation*}
    p(\pi_{j_1j_2}^{(s)} \mid \bm G^{(s)}) = 
    \textrm{Beta}
    (\alpha  + \sum_{k_1, k_2} g^{(s)}_{j_1k_1, j_2k_2}, \beta + K^2 -  \sum_{k_1, k_2} g^{(s)}_{j_1k_1, j_2k_2}).
\end{equation*}
Apparently, a prior with $\alpha = 2$ and $\beta = 7$ is dominated by  $K^2 = 100$. Therefore, we raise the values of $\alpha$ and $\beta$ so that they are comparable to the magnitude of $K^2$. The graphs with $\alpha = 10$ and $\beta = 40$ are indeed denser and have better explainability. We re-ran the simulation study with $\alpha = 10$ and $\beta = 40$. The  results are in Table~\ref{table-performance-fgm-app} and  are almost identical to the previous results with $\alpha = 2$ and $\beta = 7$.

\begin{table}[H]
{
\centering
\begin{tabular}{|c| cc|c|}  
\hline
&  $s=1$ & $s=2$\\ 
 \hline
 TPR &  0.72 (0.1) &  0.86(0.07)\\
FPR & 0.05 (0.02) & 0.09 (0.02)\\
MCC &  0.65 (0.1) &  0.69 (0.1)\\
\hline
\end{tabular}               
\caption{True positive rate (TPR), false positive rate (FPR), and Matthews correlation coefficient (MCC) for graph estimation from the simulation study with $\alpha = 10$ and $\beta = 40$. }
\label{table-performance-fgm-app}
}
\end{table} 

\begin{figure} 
\centering
\begin{tabular}{cccc}
\includegraphics[width=0.4\textwidth]{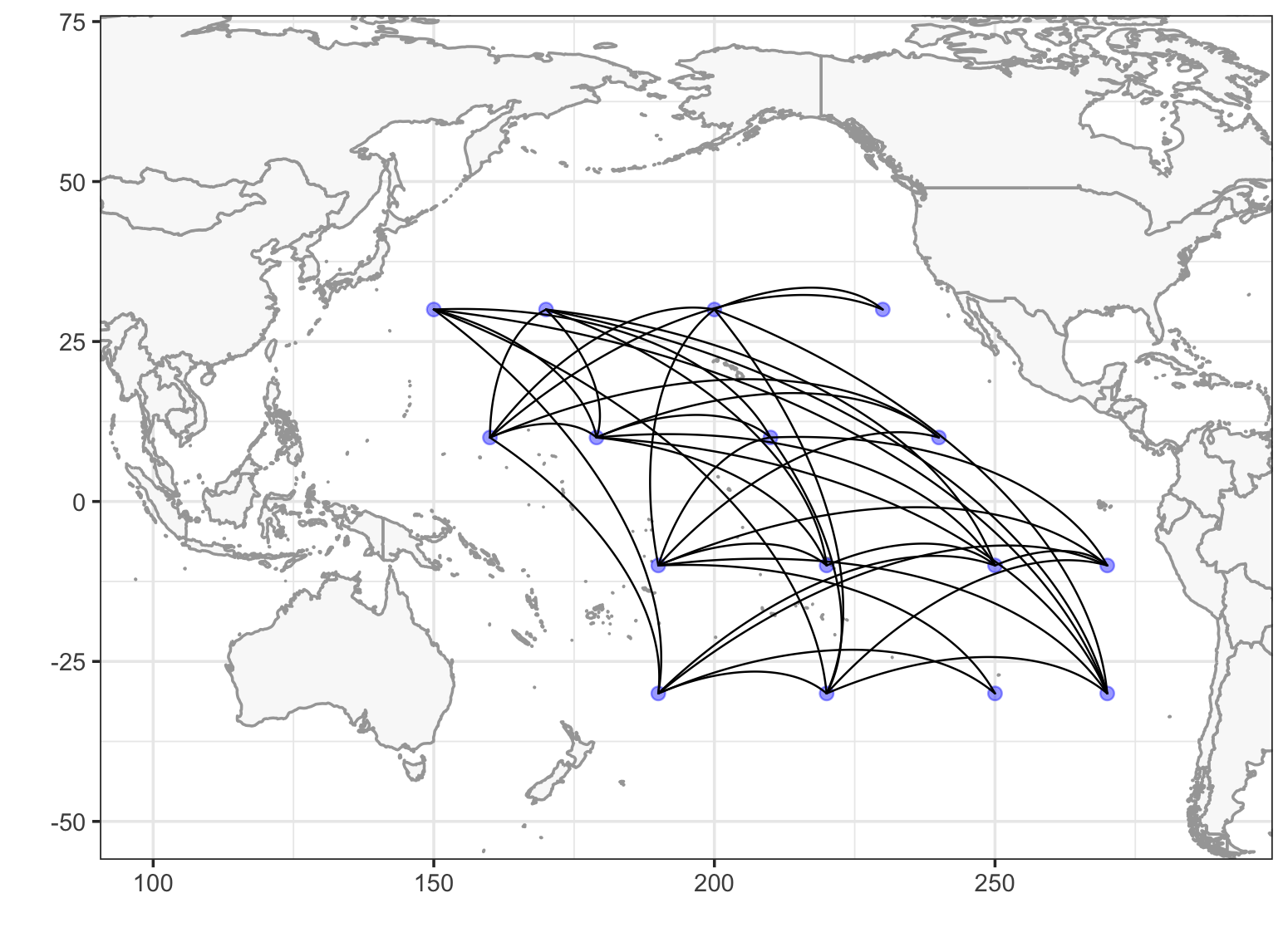}&
\includegraphics[width=0.4\textwidth]{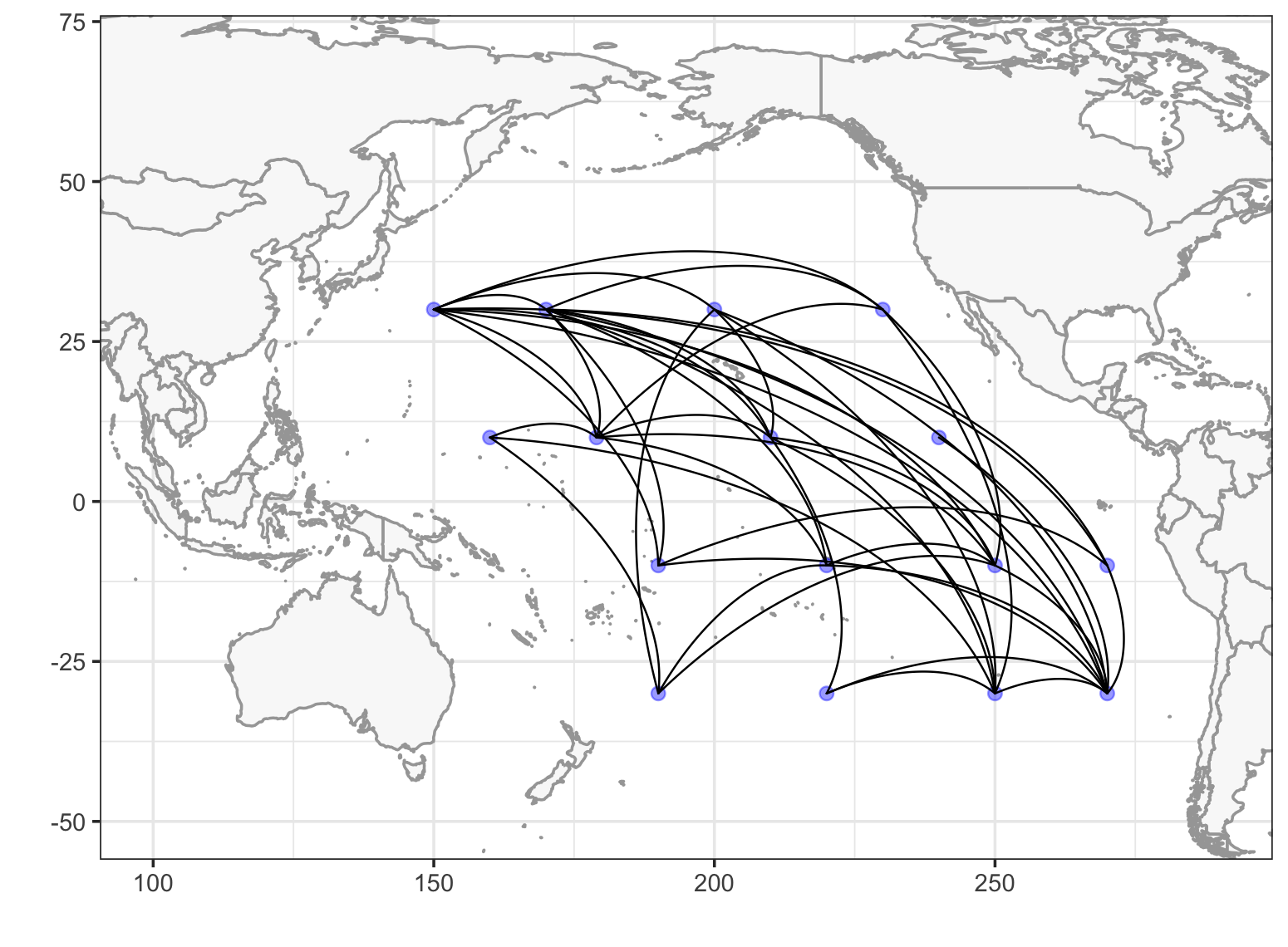} \\
\includegraphics[width=0.4\textwidth]{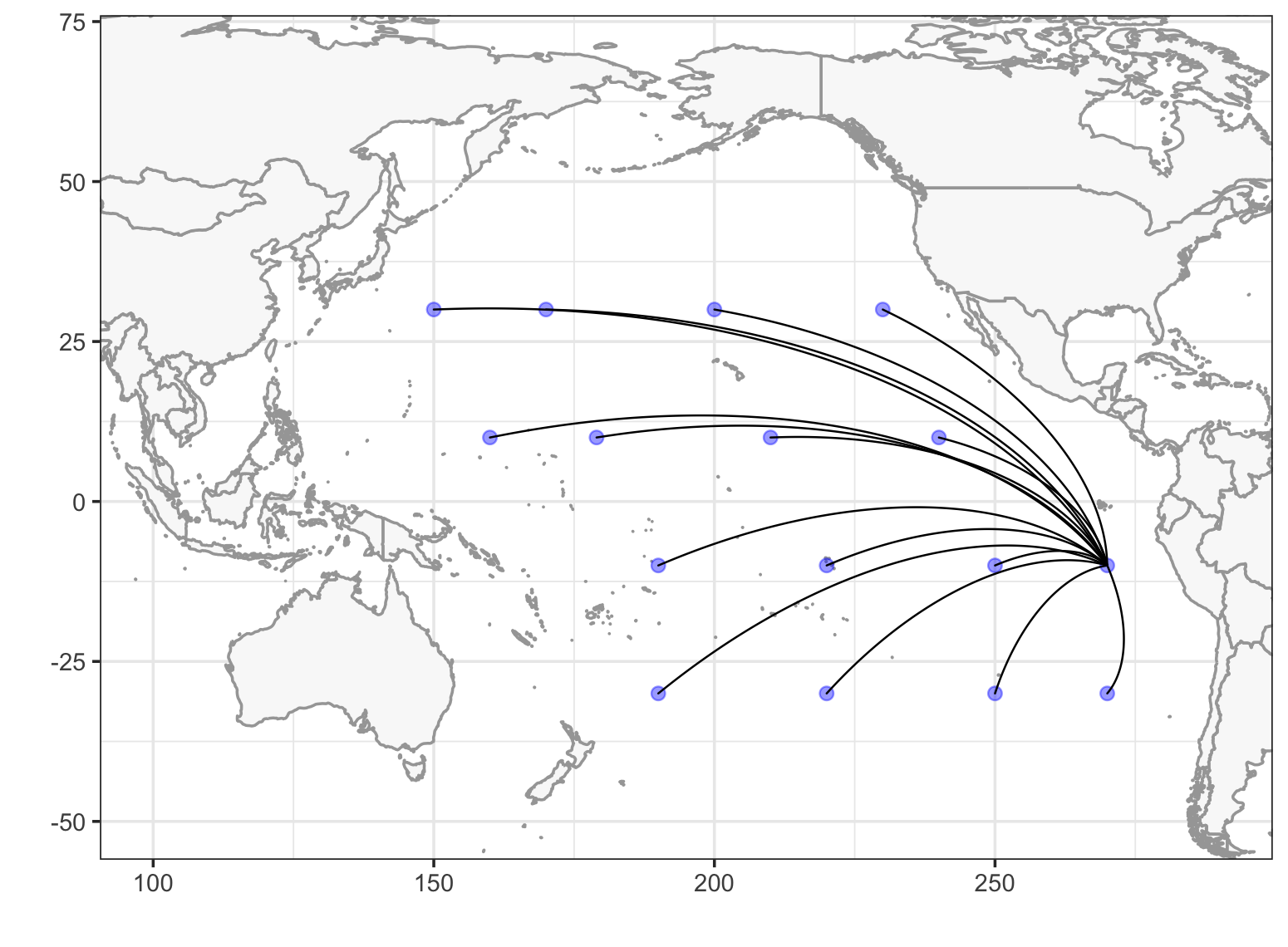} &
\includegraphics[width=0.4\textwidth]{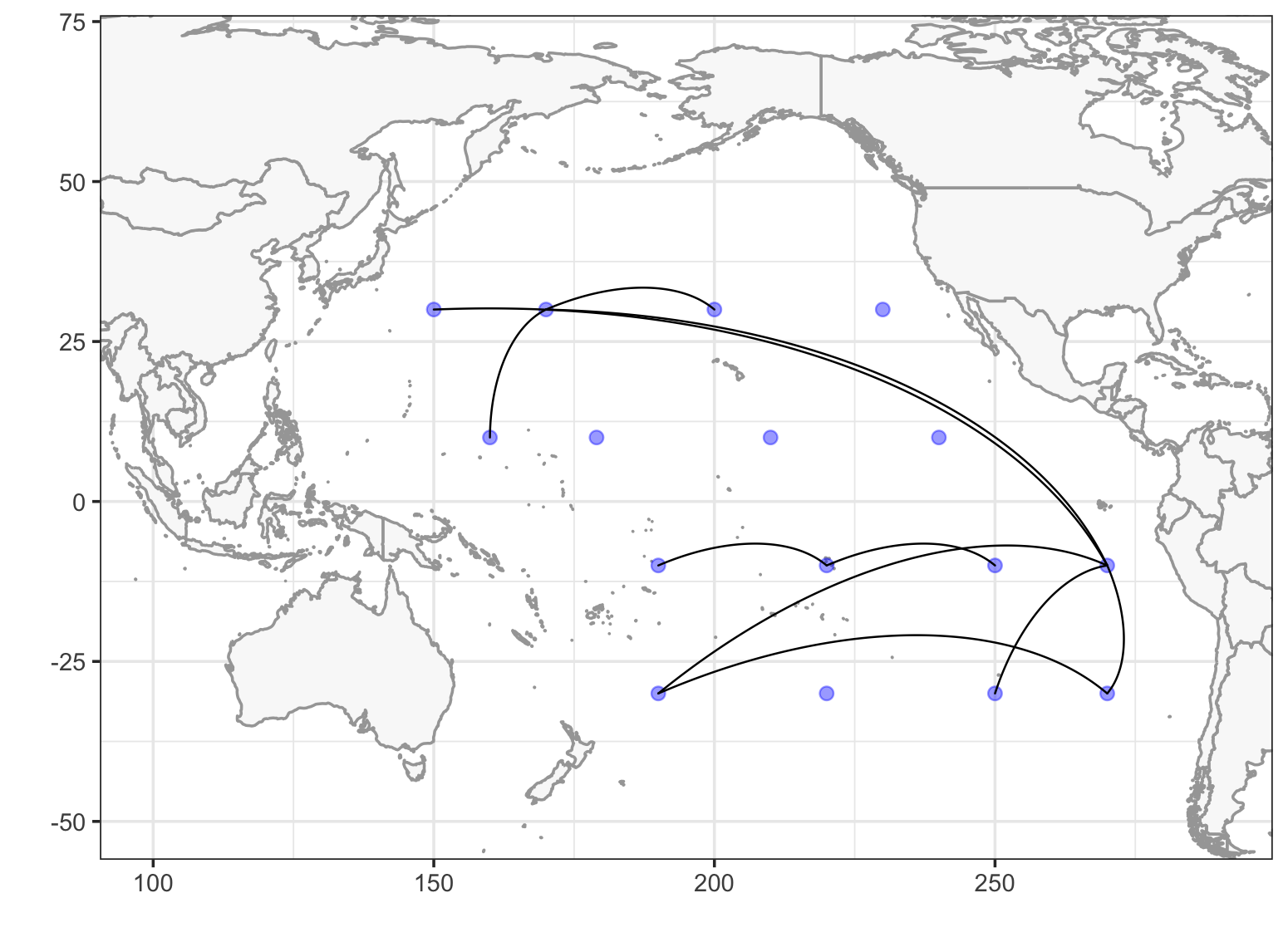}
\end{tabular}
\caption{Posterior estimation of 4 graphs in spring (top left), summer (top right), fall (bottom left) and winter (bottom right).}
\label{sst-four-season}
\end{figure}

In addition, we fit a three-changepoint model in order to find patterns in the four seasons. The changepoints are constrained inside 3 intervals so that each segment is about 3-month long. Posterior graph estimations are plotted in Figure \ref{sst-four-season}. The connections are consistent with those found when $S = 2$. For example, there are more zonal connections on the Western side from March to August, and the graph is dominated by teleconnections from the North West to the South East from September to February. The DIC of this 3-changepoint model is -68887. In conclusion, it is reasonable to combine four seasons into two segments.

\end{appendices}

\end{document}